\documentclass[%
nofootinbib,
reprint,
longbibliography,
superscriptaddress,
]{revtex4-1}

\makeatletter
\newcommand{\fmarki}{*}
\newcommand{\fmarkii}{\ensuremath{\dagger}}
\newcommand{\fmarkiii}{\ensuremath{\ddagger}}
\newcommand{\fmarkiv}{\ensuremath{\mathsection}}
\newcommand{\fmarkv}{\ensuremath{\mathparagraph}}
\newcommand{\fmarkvi}{\ensuremath{\|}}
\newcommand{\fmarkvii}{**}
\newcommand{\fmarkviii}{\ensuremath{\dagger\dagger}}
\newcommand{\fmarkix}{\ensuremath{\ddagger\ddagger}}
                
\def\@fnsymbol#1{{\ifcase#1\or \fmarki\or \fmarkii\or \fmarkiii\or \fmarkiv\or \fmarkv\or \fmarkvi\or \fmarkvii\or \fmarkviii\or \fmarkix \else\@ctrerr\fi}}
\makeatother

\renewcommand{\fmarki}{$\dagger$}
\renewcommand{\fmarkii}{$\ddagger$}
\renewcommand{\fmarkiii}{$\mathsection$}

\usepackage[dvipdfmx]{graphicx}
\usepackage{graphicx}
\usepackage{amsmath,amssymb,amsthm,mathrsfs,amsfonts,dsfont}
\usepackage{subfigure, epsfig}
\usepackage{bm}
\usepackage{enumerate}
\usepackage{comment}
\usepackage{mathtools}
\usepackage{tikz}

\usepackage[colorlinks,linkcolor=blue,citecolor=blue]{hyperref}
\usepackage[capitalise]{cleveref}

\usepackage{qcircuit}

\usepackage{perpage}
\MakePerPage{footnote}

\usepackage[normalem]{ulem} 

\newcommand{\ket}[1]{|{#1}\rangle}
\newcommand{\bra}[1]{\langle{#1}|}

\begin{document}

\title{Diagonalization of large many-body Hamiltonians on a quantum processor}

\author{Nobuyuki Yoshioka$^*$}
\email{nyoshioka@ap.t.u-tokyo.ac.jp}
\affiliation{Department of Applied Physics, University of Tokyo, 7-3-1 Hongo, Bunkyo-ku, Tokyo 113-8656, Japan}

\author{Mirko Amico$^*$}
\email{mamico@ibm.com}
\affiliation{IBM Quantum, T. J. Watson Research Center, Yorktown Heights, NY 10598, USA}

\author{William Kirby$^*$}
\email{william.kirby@ibm.com}
\affiliation{IBM Quantum, IBM Research Cambridge, Cambridge, MA 02142, USA}

\author{Petar Jurcevic}
\affiliation{IBM Quantum, T. J. Watson Research Center, Yorktown Heights, NY 10598, USA}

\author{Arkopal Dutt}
\affiliation{IBM Quantum, IBM Research Cambridge, Cambridge, MA 02142, USA}

\author{Bryce Fuller}
\affiliation{IBM Quantum, T. J. Watson Research Center, Yorktown Heights, NY 10598, USA}

\author{Shelly Garion}
\affiliation{IBM Quantum, IBM Research Israel, Haifa University Campus,
Mount Carmel Haifa 3498825, Israel}

\author{Holger Haas}
\affiliation{IBM Quantum, T. J. Watson Research Center, Yorktown Heights, NY 10598, USA}

\author{Ikko Hamamura$^{**}$}
\affiliation{IBM Quantum, IBM Japan 19-21 Nihonbashi Hakozaki-cho, Chuo-ku, Tokyo, 103-8510, Japan}

\author{Alexander Ivrii}
\affiliation{IBM Quantum, IBM Research Israel, Haifa University Campus,
Mount Carmel Haifa 3498825, Israel}

\author{Ritajit Majumdar}
\affiliation{IBM Quantum, IBM India Research Lab, Bengaluru, KA 560045, India}

\author{Zlatko Minev}
\affiliation{IBM Quantum, T. J. Watson Research Center, Yorktown Heights, NY 10598, USA}

\author{Mario Motta}
\affiliation{IBM Quantum, T. J. Watson Research Center, Yorktown Heights, NY 10598, USA}

\author{Bibek Pokharel}
\affiliation{IBM Quantum, IBM Research Almaden, San Jose, CA 95120, USA}

\author{Pedro Rivero}
\affiliation{IBM Quantum, T. J. Watson Research Center, Yorktown Heights, NY 10598, USA}

\author{Kunal Sharma}
\affiliation{IBM Quantum, T. J. Watson Research Center, Yorktown Heights, NY 10598, USA}

\author{Christopher J. Wood}
\affiliation{IBM Quantum, T. J. Watson Research Center, Yorktown Heights, NY 10598, USA}

\author{Ali Javadi-Abhari}
\affiliation{IBM Quantum, T. J. Watson Research Center, Yorktown Heights, NY 10598, USA}

\author{Antonio Mezzacapo}
\affiliation{IBM Quantum, T. J. Watson Research Center, Yorktown Heights, NY 10598, USA}



\begin{abstract}
The estimation of low energies of many-body systems is a cornerstone of the computational quantum sciences. Variational quantum algorithms can be used to prepare ground states on pre-fault-tolerant quantum processors, but their lack of convergence guarantees and impractical number of cost function estimations prevent systematic scaling of experiments to large systems. Alternatives to variational approaches are needed for large-scale experiments on pre-fault-tolerant devices. 
Here, we use a superconducting quantum processor to compute eigenenergies of quantum many-body systems on two-dimensional lattices of up to 56 sites, using the Krylov quantum diagonalization algorithm, an analog of the well-known classical diagonalization technique.
We construct subspaces of the many-body Hilbert space using Trotterized unitary evolutions executed on the quantum processor, and classically diagonalize many-body interacting Hamiltonians within those subspaces.
These experiments show that quantum diagonalization algorithms are poised to complement their classical counterpart at the foundation of computational methods for quantum systems.

\end{abstract}
\maketitle

\def\thefootnote{*}\footnotetext{Co-first authors with equal contributions.}
\def\thefootnote{**}\footnotetext{Present Address: NVIDIA G.K., Tokyo 107-0052, Japan.}
\def\thefootnote{\arabic{footnote}}


Solving the Schr\"{o}dinger equation for quantum many-body systems is at the core of many computational algorithms in fields such as condensed matter physics, quantum chemistry, and high energy physics. A quantum advantage for this task would have far-reaching consequences for natural sciences.
Among approaches to using quantum computers for eigenstate calculations, two have been the primary objects of discussion to date: quantum phase estimation (QPE)~\cite{kitaev1995phaseestimation,kitaev2002computation} including its recent advancements (e.g., Ref.~\cite{dong2022groundstate,lin2022heisenberglimited,li2023esprit}), and the variational quantum eigensolver (VQE)~\cite{peruzzo_2014}.
Experimental implementations on pre-fault-tolerant devices have focused on VQE, which has been demonstrated on various experimental platforms for a wide range of problems (e.g., Ref.~\cite{peruzzo_2014,kandala_2017,zhao2023orbitaloptimized}).
However, the bottleneck of parametric optimization has so far prevented its scaling beyond small instances.
QPE on the other hand possesses theoretical precision guarantees, but quantum error correction will be necessary to reach the circuit depths required for problems of value, although small examples have been implemented~\cite{higgins2007entanglement,lanyon2010towards,paesani2017bayesian}.

\begin{figure*}[t]
    \centering
    \begin{center}
    \includegraphics[width=0.99\linewidth]{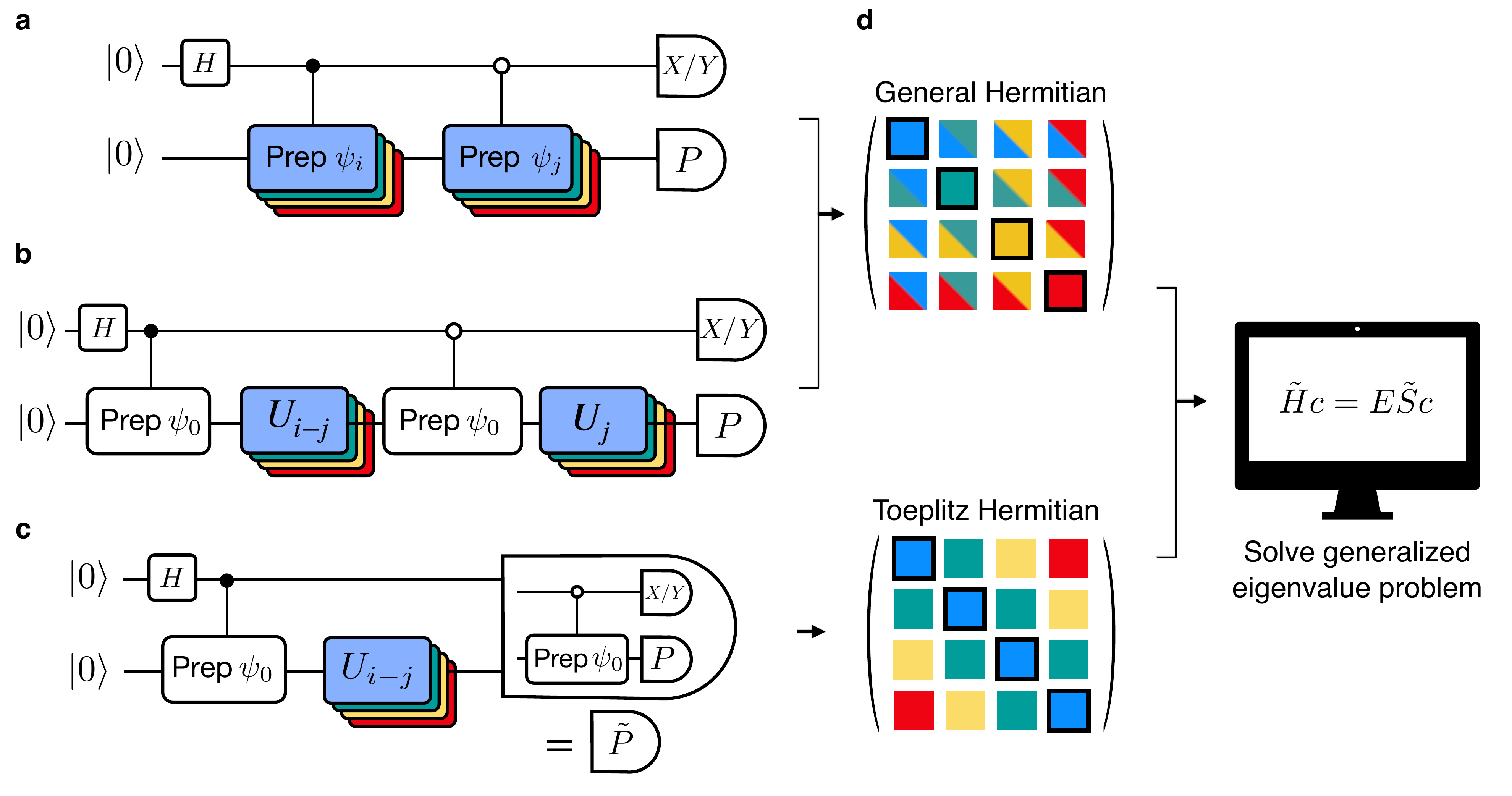}
    \caption{{\bf Schematic of Krylov quantum diagonalization.} {\bf a,} Hadamard circuit for computing matrix elements of the form $\langle \psi_i | P| \psi_j  \rangle$, which relies on controlled unitary implementation of Krylov basis states.  {\bf b,} Simplification of the circuit by exploiting a symmetry such as particle-number conservation. 
    {\bf c,}
    The construction employed in this work.  Only one time evolution circuit is required, and the second controlled preparation circuit is absorbed into the basis of the measurement.
    {\bf d,} Classical postprocessing to construct  matrices $\tilde{H}$ and $\tilde{S}$, which yield a generalized eigenvalue problem. The matrices are Hermitian for the circuits shown in {\bf a, b}, and Toeplitz Hermitian for {\bf c}. Note that the diagonal elements, enclosed by black lines, can be computed classically. 
    }\label{fig:fig1}
    \end{center}
\end{figure*}

These results leave a gap in methods for eigenstate estimation, between the small demonstrations that have been executed so far, and large-scale, high-accuracy simulations using QPE or related methods on fault-tolerant quantum computers.
In this work we demonstrate that Krylov quantum diagonalization (KQD)~\cite{parrish_2019,motta_2019,stair2020krylov,urbanek2020chemistry,cohn2021filterdiagonalization,seki2021powermethod,baker2021block,baker2021lanczos,klymko2022realtime,jamet2022greens,lee2023sampling,kirby2023exactefficient,shen2023realtimekrylov,tkachenko2024davidson,epperly2021subspacediagonalization,kirby2024analysis}, a type of quantum subspace diagonalization~\cite{mcclean_2017,colless_2018,parrish_2019,motta_2019,takeshita2020subspace,huggins2020nonorthogonal,stair2020krylov,urbanek2020chemistry,bharti2020quantum,bharti2021iterative,cohn2021filterdiagonalization,yoshioka2021virtualsubspace,epperly2021subspacediagonalization,seki2021powermethod,baker2021block,baker2021lanczos,cortes2022krylov,klymko2022realtime,jamet2022greens,baek2023nonorthogonal,lee2023sampling,zhang2023measurementefficient,kirby2023exactefficient,shen2023realtimekrylov,kirby2024analysis,yang2023shadow,ohkura2023leveraging,tkachenko2024davidson,motta2024subspace}, can fill the gap for more general problems.

The main idea in KQD is to use the quantum computer to approximate the projection of the Hamiltonian into a Krylov space spanned by various time evolutions of an initial reference state.
The resulting low-dimensional matrix is then classically diagonalized to obtain approximate low-lying energy eigenstates~\cite{parrish_2019}.
This method shares the property of variationality with VQE (up to effects of noise), but does not require an iterative parameter optimization, instead relying on a single round of circuit executions followed by classical post-processing.
Furthermore, the accuracy of the method can be bounded theoretically~\cite{epperly2021subspacediagonalization,kirby2024analysis}, as in QPE, meaning that KQD can continue to be valuable through the transition into the fault-tolerant era.
In the near-term, time evolutions for simulations with less stringent accuracy requirements are not prohibitive for existing quantum computers.

We use KQD to estimate the ground-state energy of the Heisenberg model on a heavy-hexagonal lattice.
We show that although noise poses a significant obstacle to high accuracy even with advanced error mitigation~\cite{vandenberg2023probabilistic,kim2023evidence}, we can obtain convergence to the ground-state energy on up to 56 qubits.

KQD consists of two main steps.
The first is a quantum subroutine to construct the matrices
\begin{eqnarray}
\label{eq:general_matrix_elements}
    \widetilde{H}_{jk} \coloneqq \langle \psi_j | H | \psi_k \rangle,\qquad\widetilde{S}_{jk} \coloneqq \langle \psi_j | \psi_k \rangle,
\end{eqnarray}
which correspond to the projection of the Hamiltonian into and the overlap (Gram) matrix of a subspace ${\mathcal{K}\coloneqq {\rm Span}\{|\psi_0\rangle, ..., |\psi_{D-1} \rangle\}}$.
The second step is to classically solve the time-independent Schr\"{o}dinger equation projected into the subspace, which is given by
\begin{eqnarray}
\label{eq:gen_eig_prob}
    \widetilde{H}c &=& E\widetilde{S}c,
\end{eqnarray}
where $c$ is a coordinate vector in the Krylov space.
The approximate ground-state energy, within the entire Hilbert space or a symmetry sector, is obtained as the lowest eigenvalue of \eqref{eq:gen_eig_prob}.
Two distinct components affect the accuracy of the approximation~\cite{epperly2021subspacediagonalization,kirby2024analysis}: the intrinsic error of projecting the full eigenvalue problem down into the subspace, which is related to the overlap of sufficiently low-energy states with the subspace, and any additional algorithmic, statistical, and hardware errors.

Subspace diagonalization methods differ primarily in the choice of subspace.
In classical computing, one of the common approaches is to construct the subspace by generating correlation via local operators such as the hopping terms for fermions as in multi-reference configuration interaction~\cite{yoshioka2021solving}.
Alternatively, one can use global operators. 
For instance, the classical Lanczos method employs the power series of the Hamiltonian to construct the subspace as $\mathcal{K}_{P}={\rm Span}\{H^j |\psi_0\rangle\}$, which is also referred to as the power or polynomial Krylov space.
The main advantage of such a construction is that the accuracy of the solution improves exponentially with the subspace size $D$~\cite{kaniel1966linearalgebra,paige1971computation,saad1980lanczos}.
The limiting factor in classical Lanczos and related methods is that they inevitably suffer memory consumption that grows exponentially with the system size, owing to the need to represent entangled quantum states.

While various adaptations of this scheme to quantum computers have been proposed~\cite{mcclean_2017,colless_2018,parrish_2019,motta_2019,takeshita2020subspace,huggins2020nonorthogonal,stair2020krylov,urbanek2020chemistry,cohn2021filterdiagonalization,yoshioka2021virtualsubspace,epperly2021subspacediagonalization,seki2021powermethod,cortes2022krylov,klymko2022realtime,jamet2022greens,baek2023nonorthogonal,lee2023sampling,zhang2023measurementefficient,kirby2023exactefficient,shen2023realtimekrylov,kirby2024analysis,yang2023shadow,ohkura2023leveraging,tkachenko2024davidson,motta2024subspace}, the most appropriate for near-term quantum computers is to use real-time evolutions as the global operators to generate the Krylov space:
\begin{equation}
\label{eq:real_time_krylov}
    \mathcal{K}_U={\rm Span}\{U^j |\psi_0\rangle\}, \quad j=0,1,...,D-1,
\end{equation}
where $U\coloneqq e^{-iH\,dt}$ is the time evolution operator for some timestep $dt$~\cite{parrish_2019,motta_2019,stair2020krylov,urbanek2020chemistry,cohn2021filterdiagonalization,epperly2021subspacediagonalization,seki2021powermethod,klymko2022realtime,jamet2022greens,lee2023sampling,kirby2023exactefficient,shen2023realtimekrylov,kirby2024analysis,tkachenko2024davidson}.
The advantage of this is two-fold: first, time evolutions can be approximated by circuits of short enough depth to be implemented on existing quantum devices.
Second, one can show that even in the presence of noise, the error due to projection into this unitary Krylov space converges exponentially quickly with the Krylov dimension, just as in classical Krylov algorithms. 
The noise simply contributes an additional error term as long as it is not so large that it completely overwhelms the signal~\cite{epperly2021subspacediagonalization,kirby2024analysis}.
This means that it is possible to reach convergence of the approximate ground-state energy with a Krylov space of limited dimension.

While evaluation of the Krylov matrices on the quantum computer resolves the issue of memory, which is the main obstacle to scaling on the classical side, the main obstacle on the quantum side is noise.
Two major contributions are statistical noise due to finite shot sampling, and hardware noise in the device.
Algorithmic error from the approximation of time evolutions also enters, but below we show numerically that its effects are below the level of the hardware errors.
On the other hand, suppressing and mitigating those hardware errors proves to be crucial in order to scale the size of the simulation: we apply experimental techniques for this purpose (see Sec.~IV in the Supplemental Information for details) as well as keeping the quantum circuit as shallow as possible while maintaining global coupling structure of the Krylov space.

To simplify our circuits, we exploit the U(1) symmetry possessed by many condensed matter models including the Heisenberg model we focus on.
As a qubit operator, U(1) symmetry can be expressed as conservation of Hamming weight; in terms of spin-1/2 operators it corresponds to conservation of the $z$ component of total spin.
Equivalently, we can think of the symmetry subspaces as {\it $k$-particle subspaces}, treating $\uparrow(\downarrow)$ spins as absence (presence) of a particle.

Figure~\ref{fig:fig1} shows a sequence of circuits that could in principle be used to calculate the matrix elements \eqref{eq:general_matrix_elements}.
Panel (a) shows the standard Hadamard test, which would be the default tool for such a calculation.
Panel (b) illustrates how we use spin conservation to avoid implementing the controlled time evolutions present in the conventional Hadamard test: instead, we implement controlled initializations of the reference state $|\psi_0\rangle$, and then rely on the fact that the time evolutions preserve the ``vacuum state'' $|00...0\rangle$ up to a classically-calculable phase.

As a second simplification, we note that for the exact time evolutions, $\langle\psi_0|U_j^\dagger HU_k|\psi_0\rangle=\langle\psi_0|HU_{k-j}|\psi_0\rangle$, which gives us two formally equivalent ways to measure the same matrix element, with the second yielding a simpler circuit since it only involves one time evolution.
However, once the time evolutions are approximated by Trotterization, these two expressions are no longer equal.
In Fig.~\ref{fig:fig1}c, we show the circuit corresponding to the latter version.

\begin{figure}[tb]
    \centering
    \begin{center}
    \includegraphics[width=0.75\linewidth]{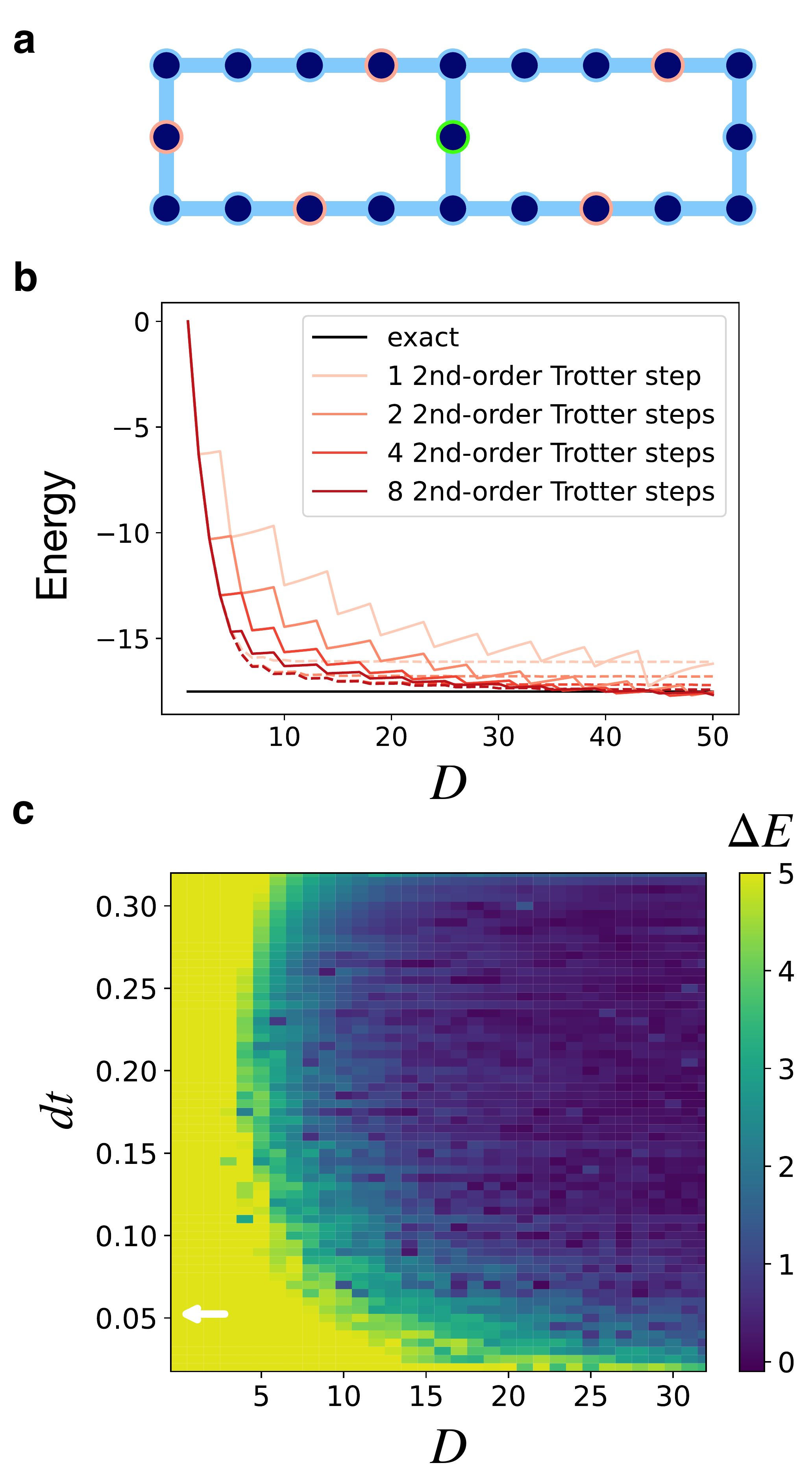}
    \caption{{\bf Numerical investigations of algorithmic errors.} {\bf a,} (20+1)-qubit layout of the Heisenberg model used for numerical simulations, with the green and red circles indicating the control and excited qubits.  {\bf b,} Energy versus Krylov space dimension. The dotted and solid lines indicate the results from the circuits in Figs.~\ref{fig:fig1}b and \ref{fig:fig1}c, respectively.
    {\bf c,} Heat map of the ground-state energy error $\Delta E$ for $k=5$-particle sector with various $dt$ and $D,$ using 4 second-order Trotter steps. The white arrow indicates the value of $\pi/\|H\|$.\label{fig:numerics}}
    \end{center}
\end{figure}

It is not \emph{a priori} clear whether one should prefer the circuits shown in panels b or c in Fig.~\ref{fig:fig1}, purely from a Trotter error perspective.
One advantage of Fig.~\ref{fig:fig1}b is that it still corresponds to variational optimization in a subspace, since each matrix element still has the form \eqref{eq:general_matrix_elements}.
However, even this ceases to be true in the presence of finite sample and device noise~\cite{kirby2024analysis}.
Figure~\ref{fig:fig1}c, the version in which Toeplitz structure is explicitly enforced, is preferable from the perspective of circuit depth for two reasons: it only requires one time evolution, and as a result, the second controlled initialization can be applied as a Clifford transformation to the Pauli observables in the Hamiltonian rather than explicitly implemented in the circuit.
In practice, we do not see dramatic violations of variationality with this method, thanks to the regularization technique used to avoid ill-conditioning of the eigenvalue problem \eqref{eq:gen_eig_prob} (see Sec.~V in the Supplemental Information for details).
As an example of this, Figure~\ref{fig:numerics} shows examples of energy curves from exact classical simulation of a 20-qubit system, comparing the results using the circuits in Figs.~\ref{fig:fig1}b and \ref{fig:fig1}c.
These findings motivated using the version of the circuits shown in  Fig.~\ref{fig:fig1}c.

\begin{figure*}[tb]
    \centering
    \begin{center}
    \includegraphics[width=0.95\linewidth]{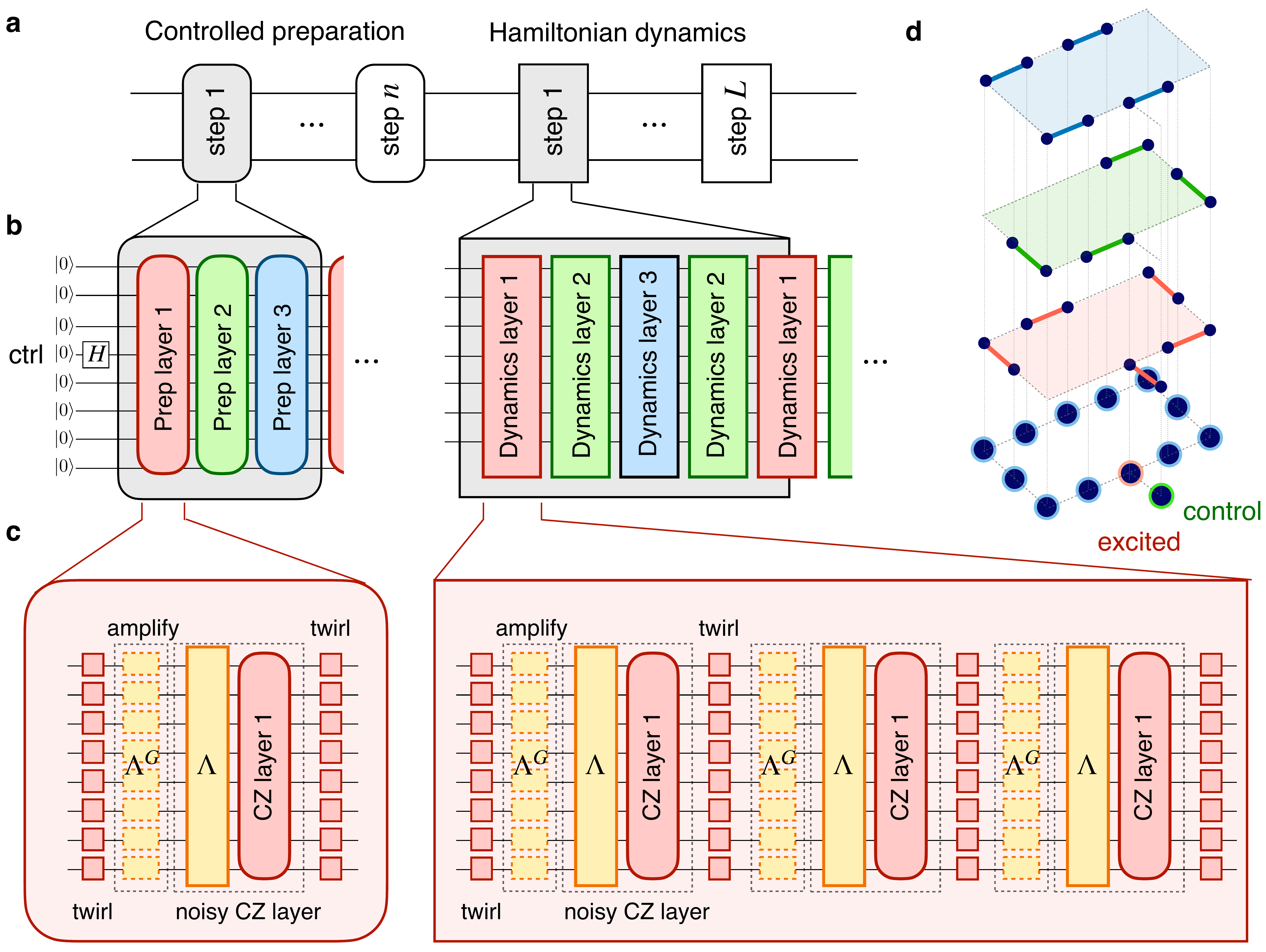}
    \caption{{\bf Quantum circuits for Krylov quantum diagonalization.} {\bf a,} Each circuit performs the controlled preparation of an initial state within the target particle sector, followed by a Trotterized time evolution. {\bf b,} The controlled preparation prepares a computational basis state in which the Hamming weight corresponds to the number of particles for the given experiment, controlled on the auxiliary qubit. Since the heavy-hex lattice can be edgewise three-colored (colors given in the figure by red, green, and blue), both the controlled preparation and the Trotterized time evolution can be implemented using sequences of three unique two-qubit gate layers interleaved with single-qubit rotations. See the main text for details. {\bf c,} Each layer of two-qubit gates is Pauli twirled in order to tailor the noise to a sparse Pauli-Lindblad noise model $\Lambda$~\cite{vandenberg2023probabilistic,kim2023evidence}, preceded by its amplification $\Lambda^G$ for PEA. Note that adjacent layers of single qubit gates, originating from either the source circuit, the twirling, or the noise amplification layer are always combined in a single layer; they are left unmerged in the figure for clarity. {\bf d,} (12+1)-qubit example of the CZ layers. 
    }\label{fig:device}
    \end{center}
\end{figure*}

\begin{figure*}[tb]
    \centering
    \begin{center}
    \includegraphics[width=0.99\textwidth]{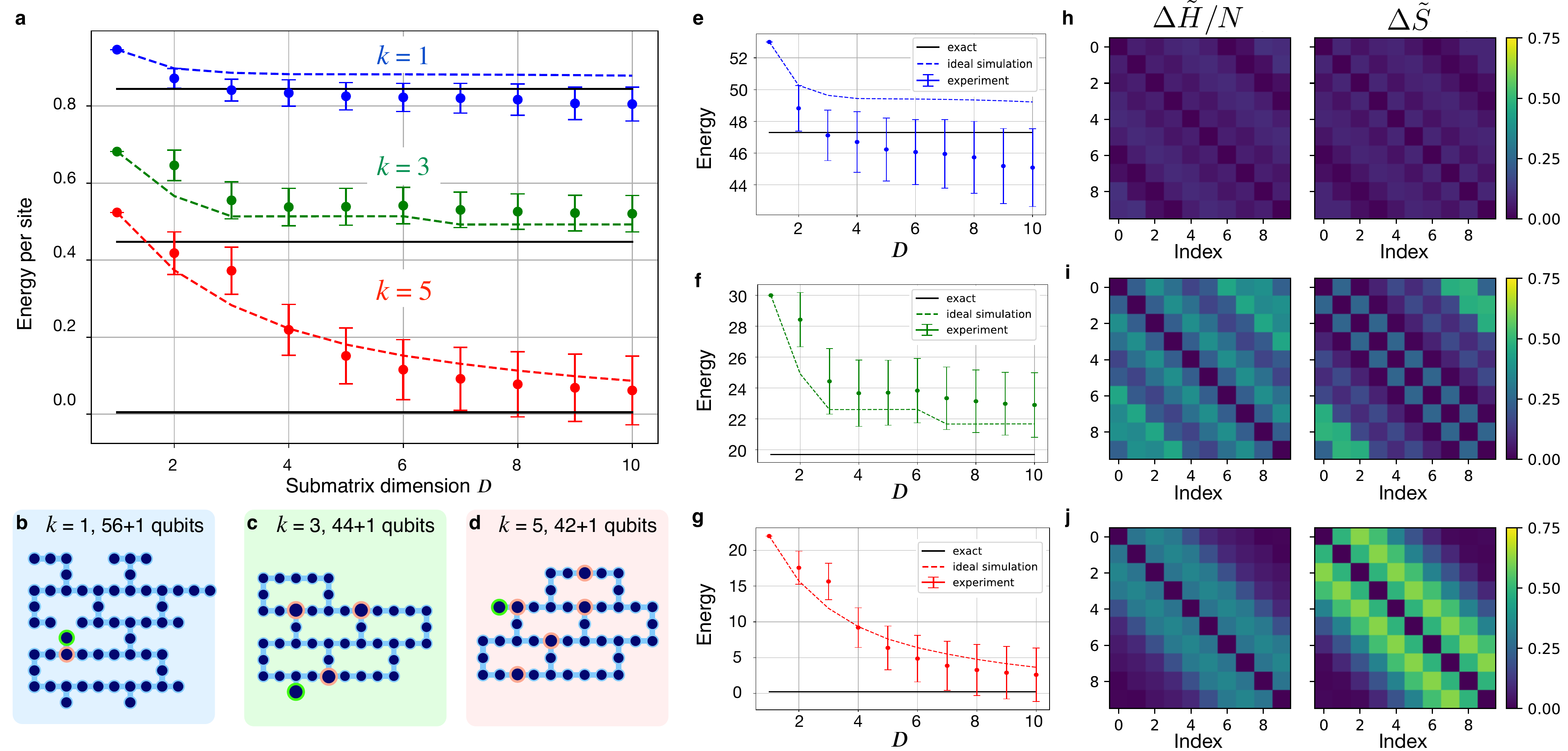}
    \caption{{\bf Experimental diagonalization of many-body Hamiltonians.} {\bf a,} The energy per site of  Heisenberg model for particle numbers $k=1,3,$ and $5$ in system sizes of $N=56, 44, $ and $42$, respectively. The error bars indicate  standard deviations estimated by bootstrapping. The dashed curves indicate the energies from noiseless classical simulations, and solid black lines show the exact lowest energy in the given $k$-particle subspace.
    {\bf b-d,} Qubit layout graphs. The green and red circles indicate the control and initial locations of particles, respectively.
    {\bf e-g,}
    Energy curves for individual particle numbers $k=1, 3$, and $5$. 
    {\bf h-j,} Error matrices $\Delta\tilde{H}/N \coloneqq |\tilde{H}_{\rm exp} - \tilde{H}_{\rm num}| /N$ and $\Delta \tilde{S} \coloneqq |\tilde{S}_{\rm exp} - \tilde{S}_{\rm num}|$, where the subscripts ``exp" and ``num" denote data from experiments and numerical calculations, respectively.
    }
    \label{fig:result_energy}
    \end{center}
\end{figure*}

For our experiments, we studied the spin-1/2 antiferromagnetic Heisenberg model, which is defined for a set of edges $E$ as
\begin{eqnarray}
\label{eq:hamiltonian}
    H = \sum_{(i, j) \in E} J_{ij} (X_i X_j + Y_i Y_j + Z_i Z_j)
\end{eqnarray}
with uniform couplings $J_{ij}=1$, where $X_i, Y_i, Z_i$ denote the Pauli matrices on the $i$th site.
The set of interactions $E$ is a subset of the heavy-hex lattice (see Fig.~\ref{fig:result_energy}).
Note that, while the heavy-hex lattice is bipartite and hence the ground state in the entire Hilbert space can be simulated efficiently using the path-integral Monte Carlo method~\cite{ceperley1995path}, the sign problem is present for excited states in general.
Among the excited states, we focus on the lowest-energy eigenstates within several $k$-particle subspaces.
The dimension of the $k$-particle subspace scales as $O(N^k)$.
Note that the circuit construction relies on the U(1) symmetry but not on SU(2) symmetry, and hence our method is directly applicable to XXZ model as well.

We ran experiments in three different $k$-particle sectors: $k=1, 3, 5$.
The initial states in all three cases were computational basis states with numbers of $|1\rangle$s given by $k$: for example, in the single-particle case, $|\psi_0\rangle=|00...1...0\rangle$.
The circuit implementations for the different values of $k$ therefore differ in the controlled preparation (see \cref{fig:fig1,fig:device}).
The $k=1$ case corresponds to generating only one particle in the initial state, which can easily be implemented with a $\mathrm{CX}$ gate between the control qubit (the ancilla in the Hadamard test) and an adjacent qubit.
For $k>1$, we chose locations for the particles that were distributed approximately uniformly over the qubit graph.

The heavy-hex lattice permits a three-coloring of its edges, in which each color corresponds to a layer of two-qubit gates that can be implemented simultaneously (see \cref{fig:device}).
Since each distinct two-qubit layer requires its own noise learning for probabilistic error amplification (PEA~--- see below), it is advantageous to minimize the number of distinct layers in the circuits.
The controlled preparation circuits can be implemented using a set of two-qubit layers corresponding to the three-coloring of edges in the heavy-hex, with only a constant overhead compared to arbitrary layers (see Sec.~II in the Supplemental Information for details).
For our Trotterized time evolutions, we partitioned the Hamiltonian terms into the same set of layers. Therefore, we only had to learn the noise models of three unique layers in total for each experiment.

The depth of the controlled-initialization part of the circuit is proportional to the distance between the two furthest apart initial particles in the qubit graph.
We used two second-order Trotter steps to approximate the time evolutions in all of our experiments.
$r$ second-order Trotter steps with three commuting groups of Hamiltonian terms requires $4r+1$ two-qubit layers (see panel b in \cref{fig:device}), yielding 9 layers in our case for the time evolution part of the circuit.

To measure the observables corresponding to real or imaginary parts of the matrix elements in $\tilde{H}$ and $\tilde{S}$, we partitioned the observables into as few locally-commuting sets (measurement bases) as possible, since such sets are co-measurable~\cite{kandala_2017}.
The shortened circuits as in the third row of Fig.~\ref{fig:fig1} require conjugating the Hamiltonian terms \eqref{eq:hamiltonian} by the second controlled-initialization circuit, since it is not physically implemented.
This yields the same number of Pauli observables since the controlled-initialization is a Clifford circuit, and one can prove that these observables can be partitioned into $2(k+2)$ measurement bases; see Sec.~II in the Supplemental Information.

We performed experiments on the Heron R1 processor \texttt{ibm\_montecarlo}.
This is a 133-qubit device with fixed-frequency transmon qubits connected to each other via tunable couplers.
Heron processors have faster two-qubit gates (similar in duration to the single-qubit gates) and lower cross-talk compared to the fixed-coupling devices of earlier generations.
To further improve the measured observables (see Fig.~\ref{fig:fig1}), we used probabilistic error amplification (PEA)~\cite{kim2023evidence} and twirled readout error extinction (TREX)~\cite{van2022model}, which mitigates SPAM errors, to approximate noise-free expectation values.
We additionally employed error suppression, in particular Pauli twirling and dynamical decoupling.
Details of the error mitigation and suppression are given in Sec.~IV of the Supplemental Information.

The size of the Krylov space was fixed to $D=10$ across all experiments. For a fixed value of $k$ the experiment was run on a specific qubit subset, chosen according to the current status of the device by using a heuristic routine for optimal qubit mapping~\cite{amico24maestro}. The $k=1$ experiment was executed on a 57-qubit subset, the $k=3$ experiment on a 45-qubit subset, and the $k=5$ experiment on a 43-qubit subset (the layouts are shown in \cref{fig:result_energy}).
The latter two were partially chosen by hand in order to have five complete heavy hexes in each case.

Although the time step $dt$ theoretically has an optimal value of $\pi/\|H\|$~\cite{epperly2021subspacediagonalization, kirby2024analysis}, the restriction to low-particle-number subspaces alters this.
Consequently, we chose the time steps heuristically, with values $0.5$, $0.022$, and $0.1$ for $k=1$, $3$, and $5$, respectively.

Results are shown in \cref{fig:result_energy}.
Panel a summarizes the results on a normalized energy scale, while e, f, and g show the convergence curves for each separate experiment.
The corresponding qubit graphs are shown in panels b, c, and d, respectively.
These convergence curves are a useful diagnostic tool for assessing the results of noisy KQD experiments.
We know from the theoretical analysis that if error rates are low enough to resolve the signal, i.e., to distinguish the lowest energy state in the Krylov space from pure noise, then we should see an exponential decay of the energy towards a value offset from the true ground-state energy by a constant depending on the error rate~\cite{epperly2021subspacediagonalization,kirby2024analysis}.
If the noise has dominated the signal, however, the rate of convergence with subspace dimension is exponentially slow with respect to system size.

In our experimental results, noise and algorithmic error (due to the Trotter approximation as well as the limited Krylov dimension) are still significant limiting factors, as evidenced by the differences between the most accurate estimated energies (at $D=10$) and the true values.
We estimated standard deviations for our experimental energies using bootstrapping, since the post-processing of solving the regularized, generalized eigenvalue problem \eqref{eq:gen_eig_prob} makes direct error propagation difficult.
This yielded the error bars in Fig.~\ref{fig:result_energy}; for further details, see Sec.~V in the Supplemental Information.
Figure~\ref{fig:result_energy} also shows the energy convergence curves for ideal classical simulations of our circuits, which are tractable by representing vectors and operators only in the restricted particle-number subspaces.
While the error bars are large due to the noisy experimental results, our estimated energies for the two larger values of $k$ are consistent with the ideal simulation curves up to these standard deviations at nearly all points.

In the $k=1$ experiment, the results deviate below the true lowest energy, indicating that noise has created an effective leakage out of the $k=1$ subspace.
This illustrates a risk of relying on symmetry conservation to remain in a particular subspace, although studying the global ground state would not be subject to this concern.
The $k=3$ and $5$ experiments were carried out later by some months, with improved device calibration, which may explain the difference.

Exact diagonalization can also be carried out in the sectors of the Hilbert space studied in the present experiments, though not in the full Hilbert space.
However, the experiments did not depend on those particular particle number sectors in any way except for the reduced circuit depth of the controlled initialization, so there are not qualitative or structural obstacles to scaling, only effects of noise.
In the specific case we focused on~--- the ground states of the Heisenberg model on a 2D heavy-hexagonal lattice~--- it is also still possible to compute precise approximations using classical techniques such as tensor networks.

The KQD approach presented here enriches the landscape of quantum algorithms for ground state estimation on pre-fault-tolerant quantum processors, filling the gap between VQE and QPE.
A complementary subspace algorithm based on sampling and sophisticated classical post-processing using a quantum-centric supercomputing (QCSC) architecture~\cite{robledomoreno2024chemistry} was recently used to demonstrate quantum simulations of chemistry beyond brute-force solutions.
This QCSC method yields classically-verifiable energies and does not require approximating time evolution, which makes it tractable in the near-term for Hamiltonians containing large numbers of terms, such as molecular Hamiltonians.
For condensed matter applications, KQD has provable convergence guarantees given an initial reference state with inverse polynomial overlap, and its circuits are feasible on pre-fault-tolerant processors as demonstrated in this work.

\noindent
{\it Acknowledgements.---}
The authors acknowledge assistance of Gadi Aleksandrowicz and Lev Bishop in the development of circuits for efficient preparation of GHZ states.
We also thank Patrick Rall, Minh Tran, Katherine Klymko, Daan Camps, Roel van Beeuman, Aaron Szasz, Yizhi Shen, Norm Tubman, Nicolas Sawaya, Giuseppe Carleo, Takahiro Sagawa, Youngseok Kim, Abhinav Kandala, Jay Gambetta, Sergey Bravyi, Hanhee Paik, and Andrew Eddins for helpful conversations.
N.Y. wishes to thank JST PRESTO No. JPMJPR2119, JST Grant Number JPMJPF2221, JST CREST Grant Number JPMJCR23I4, IBM Quantum, and JST ERATO Grant Number JPMJER2302, JST ASPIRE Grant Number JPMJAP2316.

{\it Data availability.---}
 Data  available via GitHub repository~\cite{data_github}.

 {\it Author contributions.---}
N.Y., W.K., A.D., I.H., M.M., B.P., P.R., K.S., A.J.-A., and A. M. developed the theory.
N.Y., M.A., W.K., B.F., S.G., I.H., A.I., R.M., Z.M., B.P., C.J.W., A.J.-A. wrote the codes.
M.A., P. J., H.H. ran experiments.
All authors contributed to scientific discussion and writing the paper.

\bibliography{bib}
\clearpage 
\onecolumngrid

\begin{center}
	\Large
	\textbf{Supplemental information for: Diagonalization of large many-body Hamiltonians on a quantum processor}
\end{center}

\tableofcontents

\section{Overview of Krylov quantum diagonalization}
\label{app:overview}

As described in the main text, the idea in a Krylov quantum diagonalization algorithm is to use the quantum computer to calculate the projection of the Hamiltonian $H$ into a Krylov space $\mathcal{K}$.
$\mathcal{K}$ is a Krylov space as long as it is spanned by powers of some operator applied to an initial state $|\psi_0\rangle$, but we focus more specifically on the case when the operator that generates the Krylov basis is a real time evolution, so
\begin{equation}
    |\psi_j\rangle=U_j|\psi_0\rangle=e^{-ijH\,dt}|\psi_0\rangle,\quad j=0,1,...,D-1.
\end{equation}
This time evolution can then be approximated by a quantum circuit, and the matrix elements representing the projection into the Krylov space can be calculated as in Eq.~(1) in the main text, which we reproduce here for convenience:
\begin{eqnarray}
\label{eq:general_matrix_elements_app}
    \widetilde{H}_{jk} \coloneqq \langle \psi_j | H | \psi_k \rangle,\qquad\widetilde{S}_{jk} \coloneqq \langle \psi_j | \psi_k \rangle.
\end{eqnarray}
A convenient shorthand is to represent the Krylov basis as a matrix $V$ whose columns are the vectors $|\psi_j\rangle$: then
\begin{equation}
    \widetilde{H}=V^\dagger H V,\qquad \widetilde{S}=V^\dagger V.
\end{equation}

The job of the classical post-processing is then to find the lowest eigenvalue of the matrix pencil $(\widetilde{H},\widetilde{S})$, which means solving the generalized eigenvalue problem Eq.~(2) in the main text, which we also reproduce here for convenience: 
\begin{equation}
\label{eq:gen_eig_prob_app}
    \widetilde{H}c=E\widetilde{S}c,
\end{equation}
where $c$ are $D$-dimensional coordinate vectors in the Krylov space.
The resulting lowest eigenvalue $\widetilde{E}_0$ approximates the true lowest eigenvalue of $H$ and its corresponding eigenvector $c_0$ yields a Ritz vector $Vc_0$ that approximates the true ground state, provided the noise is sufficiently low and the Krylov dimension $D$ is sufficiently high.

In practice the Krylov space approaches linear dependence as $D$ grows, which leads to an overlap (Gram) matrix $\widetilde{S}$ that is ill-conditioned~\cite{beckerman2017singularvalues}.
This means that $(\widetilde{H},\widetilde{S})$ must be regularized before \eqref{eq:gen_eig_prob_app} can be solved, and there are a few approaches to this, but the one we use is called \emph{eigenvalue thresholding}, which yields reasonable behavior even in the presence of noise~\cite{epperly2021subspacediagonalization,kirby2024analysis}.
The total ground state energy error using eigenvalue thresholding has a bound containing terms that depend on noise rate and Krylov dimension, with the former vanishing as the noise rate goes to zero and the latter vanishing exponentially as $D$ grows~\cite{epperly2021subspacediagonalization,kirby2024analysis}, as one would also expect from the classical analysis~\cite{paige1971computation,kaniel1966linearalgebra,saad1980lanczos}.
The method of eigenvalue thresholding is to project both $\widetilde{H}$ and $\widetilde{S}$ onto the eigenspaces of $\widetilde{S}$ whose eigenvalues are above some threshold $\epsilon>0$ before solving the resulting (in general lower-dimensional) version of \eqref{eq:gen_eig_prob_app}.
In practice, good performance is obtained by choosing $\epsilon$ proportional to the noise rate when the noise rate is known~\cite{kirby2024analysis}, but in our experiments we did not have a precise enough characterization of the effective noise rate (at the level of $\widetilde{H}$ and $\widetilde{S}$) for this to be useful.
Instead, we used a heuristic to choose as small $\epsilon$ as possible that yields exponential suppression of the energy error. See \cref{sec:automated_reg} for a detailed discussion.

\section{Algorithm details}
\label{app:circuit_derivation}

\subsection{Ideal circuit derivation}

\begin{figure}[ht!]
    \centering
    \begin{minipage}{3in}
    \Qcircuit @C=1em @R=.7em {
        & \ket{0}_a & & \gate{\hat{H}} & \ctrl{1} & \qw & \ctrlo{1} & \meter && \ustick{X\text{ or }Y} \\
        & \ket{0}^N & & \qw & \gate{\psi_0\text{-prep}} & \gate{U_{k-j}} & \gate{\psi_0\text{-prep}} & \meter & \ustick{\hspace{-0.05in}P}
    }
    \end{minipage}
    \caption{Circuit diagram for estimating the real ($X$ measurement on auxiliary qubit) or imaginary ($Y$ measurement on auxiliary qubit) part of the matrix element $\langle\psi_0|P|\psi_{k-j}\rangle=\langle\psi_0|PU_{k-j}|\psi_0\rangle$. The gate $\psi_0\text{-prep}$ represents the circuit that prepares our initial state $|\psi_0\rangle$ from $|0\rangle^N$, and $\hat{H}$ represents the Hadamard gate (not the Hamiltonian).}
    \label{fig:circ_schematic}
\end{figure}
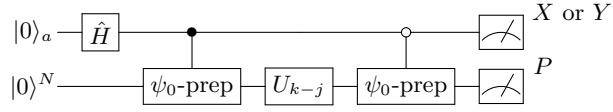

Figure~\ref{fig:circ_schematic} shows the circuit used in our experiments.
In practice, as noted in the main text, the second controlled-initialization gate is applied in the Heisenberg picture to the measured Pauli operators instead of physically on the quantum computer, but we analyze the logic of the circuit as if it were applied to the state on the quantum computer; the results are equivalent.

Assume that $|0\rangle^N$ is an eigenstate of the Hamiltonian $H$, which is satisfied when the Hamiltonian obeys U(1) symmetry represented by Hamming weight of qubit computational basis states, since in this case $|0\rangle^N$ represents the ``vacuum state.''
Further assume that we can classically efficiently calculate the eigenvalue of $|0\rangle^N$ with respect to $H$: this is always true provided $H$ has an efficient representation in terms of Pauli operators, since one can calculate the expectation value of any Pauli operator with respect to any computational basis state efficiently.
Note that this means we can calculate the phase imparted to $|0\rangle^N$ by a time evolution under $H$, and this remains true and exact even if the time evolution is Trotterized, as long as the gates in the Trotterization still exactly conserve Hamming weight.

The circuit in \cref{fig:circ_schematic} therefore implements the following state prior to measurement:
\begin{equation}
\label{transformation}
\begin{split}
    \ket{0}_a\ket{0}^N\xrightarrow{\hat{H}}&\frac{1}{\sqrt{2}}
    \left(
    \ket{0}_a\ket{0}^N+ \ket{1}_a\ket{0}^N
    \right)\\
    \xrightarrow{\text{1-ctrl-prep}}&\frac{1}{\sqrt{2}}\left(|0\rangle_a|0\rangle^N+|1\rangle_a|\psi_0\rangle\right)\\
    \xrightarrow{U_{k-j}}&\frac{1}{\sqrt{2}}\left(e^{i\phi}\ket{0}_a\ket{0}^N+\ket{1}_aU_{k-j}\ket{\psi_0}\right)\\
    \xrightarrow{\text{0-ctrl-prep}}&\frac{1}{\sqrt{2}}
    \left(
    e^{i\phi}\ket{0}_a\ket{\psi_0}
    +\ket{1}_aU_{k-j}\ket{\psi_0}
    \right)\\
    =&\frac{1}{2}
    \left(
    \ket{+}_a\left(e^{i\phi}\ket{\psi_0}+U_{k-j}\ket{\psi_0}\right)
    +\ket{-}_a\left(e^{i\phi}\ket{\psi_0}-U_{k-j}\ket{\psi_0}\right)
    \right)\\
    =&\frac{1}{2}
    \left(
    \ket{+i}_a\left(e^{i\phi}\ket{\psi_0}-iU_{k-j}\ket{\psi_0}\right)
    +\ket{-i}_a\left(e^{i\phi}\ket{\psi_0}+iU_{k-j}\ket{\psi_0}\right)
    \right)
\end{split}
\end{equation}
where we have used the phase shift $ U_{k-j}\ket{0}^N = e^{i\phi}\ket{0}$ in the third line, which is classically calculable because the eigenvalue of $\ket{0}^N$ with respect to $H$ is classically calculable, as noted above. Therefore the expectation values of the measured Pauli operators $X\otimes P$ or $Y\otimes P$ are obtained as
\begin{equation}
\begin{split}
    \langle X\otimes P\rangle&=\frac{1}{4}
    \Big(
    \left(e^{-i\phi}\bra{\psi_0}+\bra{\psi_0}U_{k-j}^\dagger\right)P\left(e^{i\phi}\ket{\psi_0}+U_{k-j}\ket{\psi_0}\right)
    \\
    &\qquad-\left(e^{-i\phi}\bra{\psi_0}-\bra{\psi_0}U_{k-j}^\dagger\right)P\left(e^{i\phi}\ket{\psi_0}-U_{k-j}\ket{\psi_0}\right)
    \Big)\\
    &=\text{Re}\left[e^{-i\phi}\bra{\psi_0}PU_{k-j}\ket{\psi_0}\right],
\end{split}
\end{equation}
\begin{equation}
\begin{split}
    \langle Y\otimes P\rangle&=\frac{1}{4}
    \Big(
    \left(e^{-i\phi}\bra{\psi_0}+i\bra{\psi_0}U_{k-j}^\dagger\right)P\left(e^{i\phi}\ket{\psi_0}-iU_{k-j}\ket{\psi_0}\right)
    \\
    &\qquad-\left(e^{-i\phi}\bra{\psi_0}-i\bra{\psi_0}U_{k-j}^\dagger\right)P\left(e^{i\phi}\ket{\psi_0}+iU_{k-j}\ket{\psi_0}\right)
    \Big)\\
    &=\text{Im}\left[e^{-i\phi}\bra{\psi_0}PU_{k-j}\ket{\psi_0}\right].
\end{split}
\end{equation}
Since $\phi$ is classically calculable, we can simply apply $e^{i\phi}$ to these measurement results to get our desired matrix elements.

Applying this scheme to each Pauli operator $P$ in the Hamiltonian allows us to build up estimates of the full Hamiltonian matrix elements
\begin{equation}
    \langle\psi_0|HU_{k-j}|\psi_0\rangle=\sum_{P}\alpha_P\langle\psi_0|PU_{k-j}|\psi_0\rangle,
\end{equation}
where $\alpha_P$ are the coefficients of the Pauli representation of the Hamiltonian ($H=\sum_{P}\alpha_PP$).
Note that because the matrix elements are calculated as expectation values of these Hamiltonian terms, we can measure more than one term at a time as long as they commute locally.
If we were applying the second controlled-initialization physically at the end of the quantum circuit, this would mean that there are only three measurement bases, since our Heisenberg Hamiltonians contain only $XX$, $YY$, and $ZZ$ terms.
The application of the second controlled-initialization to the measurement operators instead complicates this (details in Sec.~\ref{app:measurement_bases}), but we are still far better off than if we had to estimate the matrix elements of each Pauli operator one at a time.
Also, note that the matrix elements of the overlap matrix $\widetilde{S}$ are obtained by replacing $P$ with the identity, which is compatible with every measurement basis, so our estimates of matrix elements of $\widetilde{S}$ are obtained from the aggregate of all the data collected to estimate the matrix elements of $\widetilde{H}$.

\subsection{Controlled state initialization}
\label{app:initialization}

Before describing the method of constructing the controlled preparation circuits, we discuss the choice of initial state.
For $k>1$, it is important to consider the initial locations (qubit sites) of the particles.
This is because the error in the converged approximate energy output of the Krylov quantum diagonalization method depends on the overlap of the initial reference state with the true ground state (or low energy state of interest), as well as the hardware, statistical, and algorithmic noise rates~\cite{epperly2021subspacediagonalization,kirby2024analysis}.
For a challenging eigenstate approximation problem, finding a suitable initial state is handled heuristically, in our case with the additional constraint of being limited to low-depth circuits.
In particular, we confirmed by numerics on small systems the intuitive notion that an initial state whose particles are distributed roughly uniformly over the qubit graph would yield better convergence to the lowest-energy state of the subspace than a state whose particles are all adjacent.
Hence a heuristic for choosing particles in this way was used in the $k>1$ cases.

\begin{figure}[t]
    \centering
    
    \includegraphics[width=0.4\columnwidth]{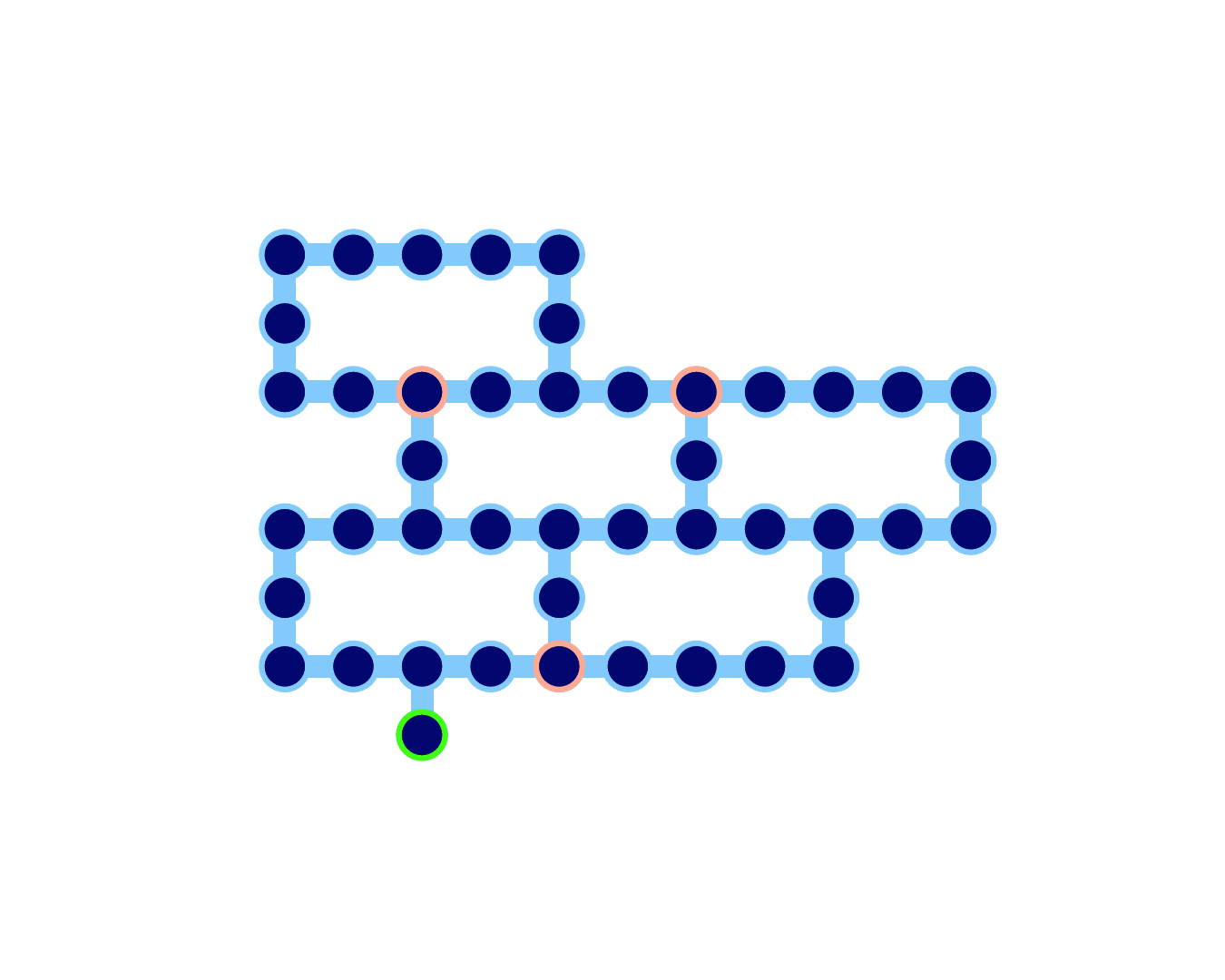}
    \includegraphics[width=0.55\columnwidth]{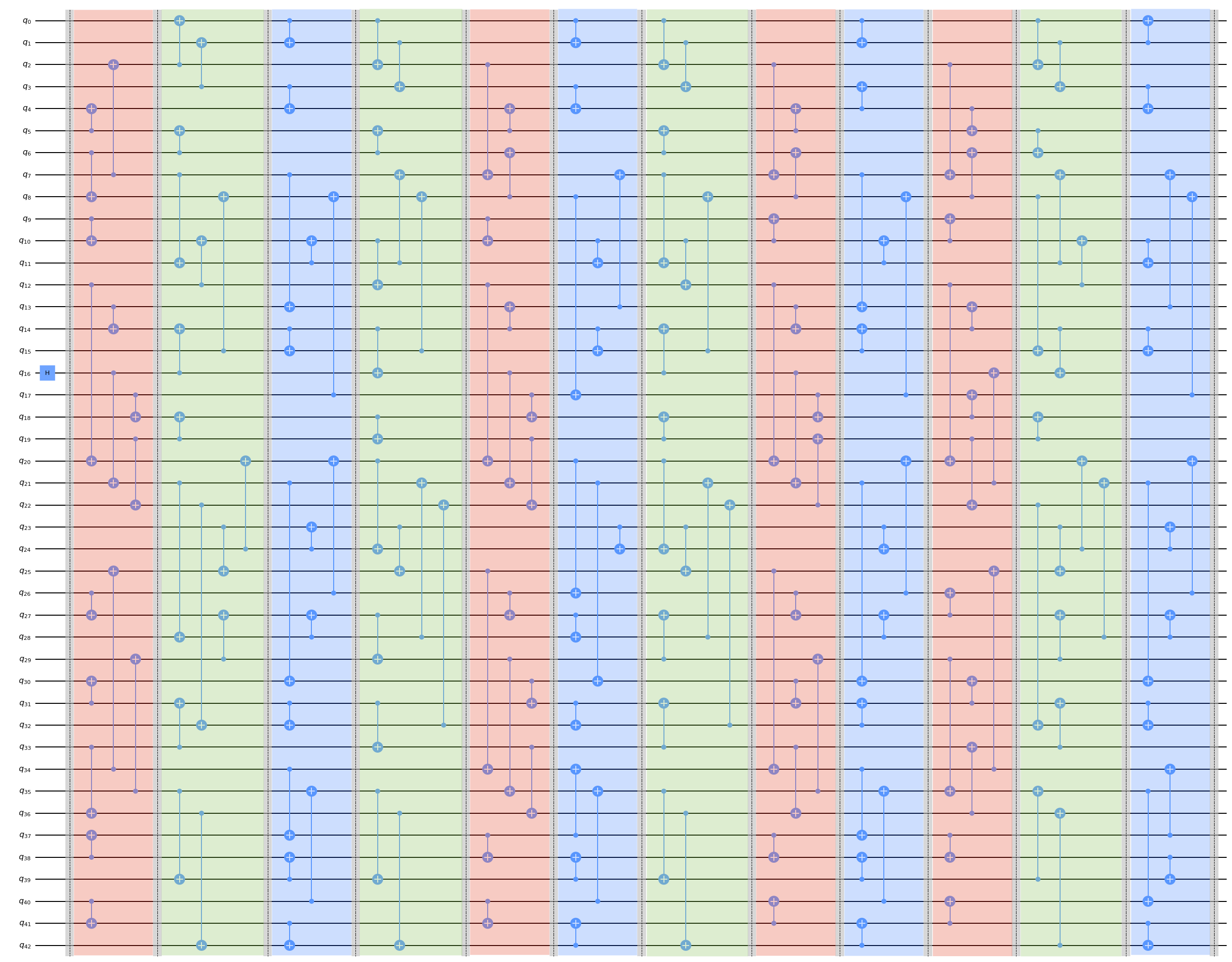}
    \caption{Controlled preparation circuit for our $k=3$ experiment (i.e., the layout and initial state shown on the left~--- see also Fig. 4c in the main text). The background colors correspond to the three distinct CX-layers, which determine the allowed simultaneous interactions up to the directions of the gates. In the layout on the left, the highlighted qubits are the locations of $1$s in the initial state, or equivalently in $\textbf{s}_\text{final}$.}
    \label{fig:k3-colored-ctrlinit}
\end{figure}

For the controlled-preparation subcircuit, we want to prepare our reference state controlled on the state of the auxiliary qubit.
Since this is preceded by a Hadamard gate on the control qubit, and the reference state is a computational basis state, this is equivalent to preparing a GHZ state on the control qubit and the qubits that are $|1\rangle$ in the reference state.
We implement this via layers of CX gates.
Let CX$_{ij}$ denote the CX gate with control $i$ and target $j$.

Throughout the controlled preparation circuit, the entire state is a superposition of two computational basis states, $\frac{1}{\sqrt{2}}\left(|0\rangle^n+|\textbf{s}\rangle\right)$, with $|\textbf{s}\rangle$ initially $|\textbf{s}_\text{init}\rangle\coloneqq|100...0\rangle$ after the Hadamard on the control qubit.
Hence any CX circuit we construct will preserve the $|0\rangle^n$ part of the superposition, only modifying the bitstring $\textbf{s}=s_1...s_n$.
The action of a CX$_{ij}$ is to transform $s_j$ to $s_i\oplus s_j$, where $\oplus$ denotes addition mod $2$.
Hence our problem can be formulated as that of mapping $\textbf{s}_\text{init}=100...0$ to $\textbf{s}_\text{final}$ (the desired reference bitstring) using as short a sequence of layers of such additions as possible, with the additions constrained to respect the device connectivity.
We furthermore have the freedom to choose any initial $\textbf{s}_\text{init}$ we like provided it contains only a single $1$, so we can optimize over this variable as well.

Let $G$ be the qubit connectivity graph.
As a warmup, suppose we did not care about employing only specific CX layers that correspond to a edge coloring of $G$.
In that case, a method for calculating the shortest depth mapping from a bitstring $\textbf{s}_\text{init}$ containing a single $1$ to $\textbf{s}_\text{final}$ could use the following algorithm:\\~\\
\textsc{Algorithm 1}
\begin{enumerate}
    \item Choose a subtree $T$ of $G$ that contains all qubits $i$ with $s_i=1$, such that all leaves in $T$ have value $1$. If a leaf in $T$ had value $0$, then it could simply be removed from $T$.
    \item Choose a root for $T$ and place $1$ at the root.
    \item Iteratively copy the $1$ outward from the root to all nodes of $T$.
    \item Correct the values of nodes in $T$ that are supposed to be $0$ by starting again from the root and propagating outward, applying CX$_{ji}$ whenever $s_i=0$, with $j$ chosen to be any child node of $i$. Since all of the leaves have value $1$, this iteration only needs to continue at most as far as the parents of leaves.
\end{enumerate}

While it might appear from this construction that the circuit's CX-depth will be twice the depth of step 3, in fact most of the gates in step 4 can be implemented in parallel with gates in step 3.
If $m$ denotes the maximum degree of any node in $T$, then in step 3 each node $i$ controls at most $m-1$ CX gates.
This means that if a CX$_{ji}$ from one of its children $j$ is required to correct $s_i$ back to $0$, that CX can be placed at most $m$ layers after the corresponding CX$_{ij}$ in step 3, which implies that the total depth is at most $m$ larger than the depth of step 3.
The worst-case CX-depth of step 3 is $m\cdot\text{depth}(T)$, so the worst-case total CX-depth is $m\big(\text{depth}(T)+1\big)$.
For the heavy-hex graph, $m=3$.

For our actual algorithm, we want to accomplish the same task, but only using a particular set of CX layers.
This both limits the CXs that we are allowed to implement in parallel, and requires that they be implemented alongside a particular set of other CXs that we might not otherwise need.
Hence the above algorithm must be modified to accomodate this.
Define a bitstring to be \emph{sparse} with respect to $G$ if for every edge $(i,j)\in G$, either $s_i=0$ or $s_j=0$ (or both).
The method is as follows:\\~\\
\textsc{Algorithm 2}
\begin{enumerate}
    \item Reduce $\textbf{s}$ to a sparse bitstring $\textbf{sp}$.
    \item Use \textsc{Algorithm 1} to construct $\textbf{sp}$, restricted to subsets of the particular set of CX layers.
    \item Fill in the missing edges in the layers coming from step 2.
    \item Map $\textbf{sp}$ to $\textbf{s}$ using the inverse of the reduction in step 1.
\end{enumerate}

Each of these steps requires explanation.
Any bitstring $\textbf{s}$ can be reduced to a sparse bitstring $\textbf{sp}$ using the specified $m$ CX-layers only once each, as follows.
For each edge $(i,j)$ in each layer, if $s_i=s_j=1$ then we can add CX$_{ij}$ to change $s_j$ to 0.
If $s_i\neq s_j$, then a CX controlled on whichever node is $0$ will have no effect, preserving the pair (since they already satisfy the sparsity constraint).
Finally, if $s_i=s_j=0$, then either direction of CX preserves the pair.
Note that the CX-layers permit CX gates in either direction on each edge in the layer.

In step 2, we implement a slightly modified \textsc{Algorithm 1} to prepare $\textbf{sp}$, using layers of CX gates that are restricted to be subsets of the allowed ``full'' CX layers.
Each layer copies $1$s from each parent node to its children, which are immediately treated as parent nodes for the purpose of all subsequent layers.
Each parent node whose value is $0$ in $\textbf{s}$ can then be reset to $0$ ``as early as possible,'' meaning by the first CX connecting it to either its parent or one of its children in a subsequent layer in the cycle.

For example, if node $i$'s parent is in layer `red' and its children are in layers `green' and `blue' (with the layers cycling in that order), then in some `red' layer $i$'s value will be set to $1$ by a CX from its parent.
$i$ will then immediately be treated as a parent, so in the following two layers `green' and `blue' it is used to set the values of its children to $1$ via CX's.
We then return to `red,' and there are a few cases to consider:
\begin{enumerate}
    \item If the value of $i$'s parent is supposed to end up $1$ in $\textbf{sp}$, then the value of $i$ in $\textbf{sp}$ has to be $0$, by the sparsity assumption.
    Hence we can use a CX from $i$'s parent to reset the value of $i$ to $0$.
    \item If the value of $i$'s parent is supposed to end up $0$ in $\textbf{sp}$, but at this point it is still $1$ (i.e., it wasn't reset by its own parent), then we use a CX from $i$ to its parent to set the parent to $0$.
    \item If the value of $i$'s parent is supposed to end up $0$ in $\textbf{sp}$, and it is already $0$ (i.e., it was reset by its own parent), then we fill in the gate for the layer by applying CX from the parent to $i$, which has no effect.
\end{enumerate}
Then, if the value of $i$ is still $1$ after the `red' layer and it is supposed to be $0$ in $\textbf{sp}$, in the `green' layer it can be reset to $0$ by its own child.
This ensures that the states of all nodes are left in their final values in $\textbf{sp}$ as early as possible, in the sense that any ``non-sparse'' pairs of adjacent $1$s are set to either $10$ or $01$ in the first layer in which they share a gate.
Once all the iterations are complete (i.e., the $1$s have been propagated all the way to the leaves), all that is left is to correct any remaining parent nodes from the final iteration whose values need to be reset to $0$.
This requires at most one additional cycle through the $m$ CX-layers.

In step 3 of \textsc{Algorithm 2}, we fill out each CX-layer from step 2 with the edges that are missing to match the prescribed complete layers.
The construction of the iteration above ensures that any pairs of qubits that do not have CX's between them already are in a sparse configuration: either $00$, $01$, or $10$.
This means that a CX that has no effect can be added for that pair, in one direction or the other.

The worst-case total depth of step 2 of \textsc{Algorithm 2} is the same as the worst-case total depth of \textsc{Algorithm 1}, since it only requires one additional cycle through the $m$ layers on top of the cycles corresponding to the depth of the tree $T$.
Step 4 then contributes one further cycle through the $m$ layers, so the total CX-depth for the controlled-preparation circuit is $m\big(\text{depth}(T)+2\big)$.

The depth of the minimum spanning tree $T$ that connects the particles (i.e., the $1$s in $\textbf{s}$) is given by $\left\lceil\frac{\text{dist}_\text{exc}}{2}\right\rceil$, where $\text{dist}_\text{exc}$ is the distance between the two farthest apart particles.
Hence, inserting $m=3$ for the heavy-hex graph, the total depth of the controlled-preparation circuit is $3\big(\left\lceil\frac{\text{dist}_\text{exc}}{2}\right\rceil+2\big)$.

An example is shown in \cref{fig:k3-colored-ctrlinit}.
As can be see from the example, some further optimization is possible on top of the construction described above, but this is specific to the particular subgraph and locations of particles, so we will omit these details.

\subsection{Time evolution}

Time evolutions under arbitrary Hamiltonians cannot be simulated exactly on quantum computers, but they can be approximated with controllable accuracy.
In the present work, we used a Trotter approximation of the time evolutions~\cite{suzuki1985decomposition,huyghebaert1990product,suzuki1991product,lloyd1996quantumsimulators,berry2007sparse,thalhammer2008highorder,clark2009TFIM,whitfield2011quantumsimulation,kliesch2014liebrobinson,babbush2015chemical,wecker2015solving,somma2016trotter,hadfield2018divide,childs2019fasterquantum,childs2019toward,childs2021trotter}, which means partitioning the Hamiltonian into commuting subsets of terms whose evolutions can be simulated exactly, then combining sequences of those term-by-term evolutions to approximate the full evolution.

Our Hamiltonian is given by Eq.~(4) in the main text, which we reproduce here for convenience:
\begin{eqnarray}
\label{eq:hamiltonian_app}
    H = \sum_{(i, j) \in E} J_{ij} (X_i X_j + Y_i Y_j + Z_i Z_j).
\end{eqnarray}
Hence, one natural choice of commuting subsets would be to group the $XX$, $YY$, and $ZZ$ terms.
However, this is not optimal for circuit implementation since even though for example all of the $XX$ terms commute, they can overlap qubitwise, so would still need to be implemented sequentially.

\begin{figure}[t]
    \centering
    \includegraphics[width=0.5\columnwidth]{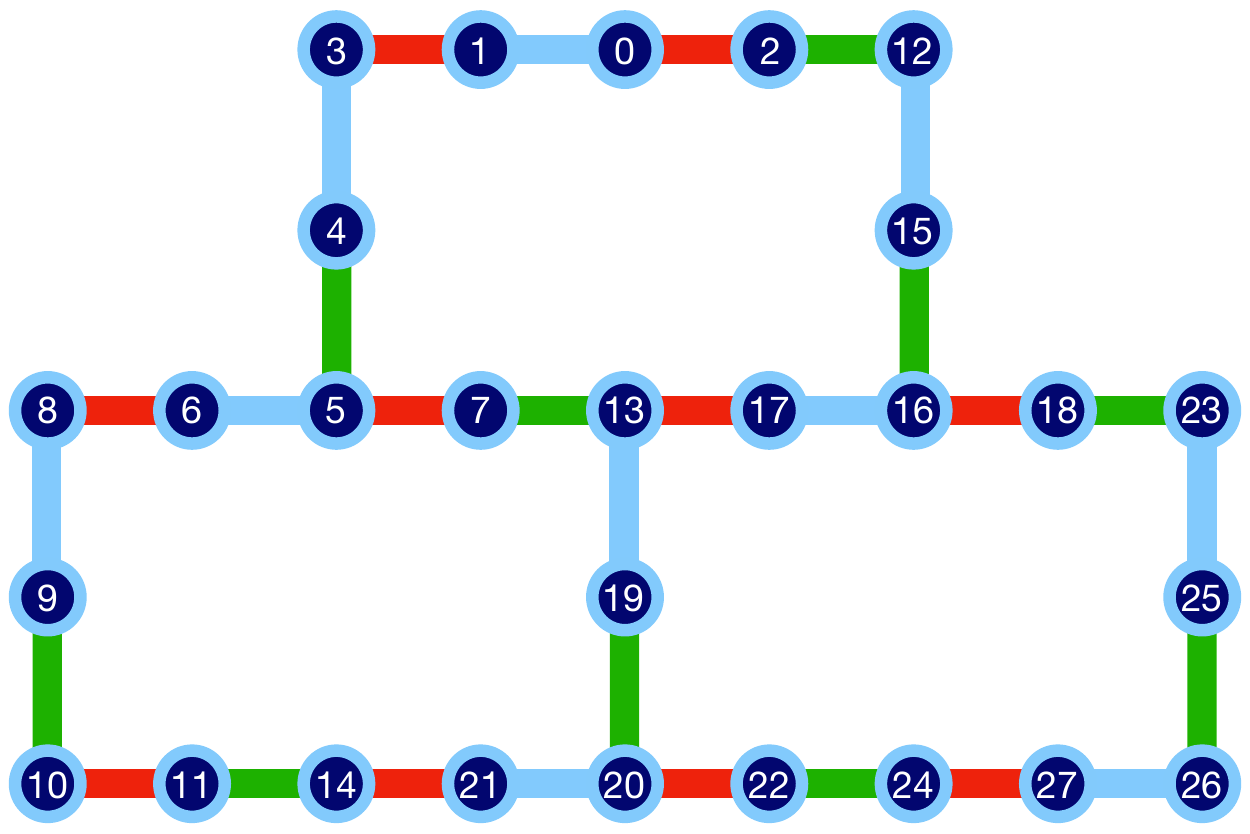}
    \caption{Example of edge three-coloring on a heavy-hex lattice.}
    \label{fig:edge_coloring_example}
\end{figure}

A choice of partition that avoids this is to color the edges $E$ in the qubit interaction graph, and let each color correspond to a commuting subgroup.
As discussed in the main text and in further detail below in \cref{app:err_supp}, all of our qubit interaction graphs are subsets of a heavy-hex graph, which admits an edge three-coloring, i.e., the edges in $E$ can be partitioned into three groups (that we label `red,' `green,' and `blue') such that edges in the same group are nonintersecting.
An example is given in \cref{fig:edge_coloring_example}.
Each edge evolution for an edge $(i,j)$ requires a rotation generated by $J_{ij}(X_iX_j+Y_iY_j+Z_iZ_j)$, and we can use arbitrary two-qubit unitary compilation to implement any such rotation with a two-qubit primitive gate depth ($\mathrm{CX}$ or $\mathrm{CZ}$ depend on the hardware backend) of three.

\begin{figure}[t]
    \centering
    \includegraphics[width=\columnwidth]{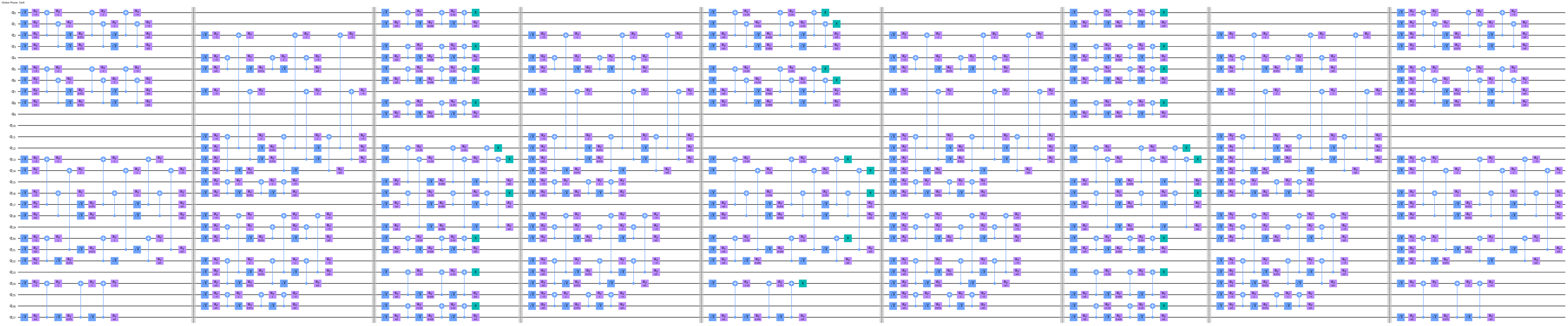}
    \caption{Example of the Trotter circuit corresponding to the qubit layout in \cref{fig:edge_coloring_example}, showing the decomposition into $\mathrm{CX}$ and single-qubit gates. The details of the single-qubit gates are unimportant: note however that each layer (separated by barriers) has a $\mathrm{CX}$-depth of three (nonoverlapping $\mathrm{CX}$ gates are implemented in parallel), and there are nine layers, as in \eqref{eq:trotter_circ}.}
    \label{fig:example_trotter_circ}
\end{figure}

Having chosen a partition of the Hamiltonian into groups of terms that can be implemented as layers of simultaneous two-qubit gates, we finally need to choose the particular Trotter approximation.
This involves balancing the accuracy of the approximation against the depth of the circuit, which is a direct tradeoff since higher depth generally permits higher accuracy Trotter approximations, but also leads to more severe device error accumulation.
The latter was sufficiently severe that we kept the depth relatively low, choosing to implement our approximate time evolutions as three second-order Trotter steps for the $k=1$ experiment and two second-order Trotter steps for $k=3$ and $5$.
If we let $R$, $G$, and $B$ denote our three groups of terms, then for example two second-order Trotter steps means the following, as a sequence of unitary matrices:
\begin{equation}
\label{eq:trotter_circ}
\begin{split}
    &\text{Trotter circuit}=\prod_{j=1}^2\left(\prod_{i\in R}U_i(t/4)\prod_{i\in G}U_i(t/4)\prod_{i\in B}U_i(t/2)\prod_{i\in G}U_i(t/4)\prod_{i\in R}U_i(t/4)\right)\\
    &~=\prod_{i\in R}U_i(t/4)\prod_{i\in G}U_i(t/4)\prod_{i\in B}U_i(t/2)\prod_{i\in G}U_i(t/4)\prod_{i\in R}U_i(t/2)\prod_{i\in G}U_i(t/4)\prod_{i\in B}U_i(t/2)\prod_{i\in G}U_i(t/4)\prod_{i\in R}U_i(t/4),
\end{split}
\end{equation}
where in the first line, the product inside the parentheses is a single second-order Trotter step.
There are two advantages to using second-order Trotter steps compared to first-order Trotter steps: first, although the latter superficially appear preferable in terms of depth, they are correspondingly worse approximations, which means more steps are required to reach the same accuracy.
Second, a second-order Trotter step has lower two-qubit gate depth than two first-order steps, since as shown in the first line in \eqref{eq:trotter_circ}, the middle layer of gates is not implemented twice but instead just has double the gate angle.
When more than one second-order Trotter step is implemented in sequence, the improvement becomes even more pronounced, since the first and last layers are the same in each step, so the first layer of each step (other than the first step) can be combined with the last layer of the previous step, which reduces the two-qubit gate depth.
This is illustrated in the second line in \eqref{eq:trotter_circ}; in general, the two-qubit gate depth of $n$ second-order Trotter steps (when there are three layers of terms in the Hamiltonian) is $3(4n+1)$, as opposed to the two-qubit gate depth of $2n$ first-order Trotter steps, $3(6n)$.
The Trotter circuit corresponding to \eqref{eq:trotter_circ} for the example qubit layout in \cref{fig:edge_coloring_example} is given in \cref{fig:example_trotter_circ}.

A remaining choice is the size of the timestep $dt$.
The error analysis in~\cite{epperly2021subspacediagonalization} showed that a sufficiently small timestep is $\pi/\|H\|$, and that it is preferable up to a point to underestimate this value rather than overestimate, since overestimating can allow contributions from high-energy states to corrupt even the optimal state in the Krylov space.
On the other hand, choosing $dt$ to be too small leads to worse conditioning of the Krylov subspace, since the Krylov basis vectors differ less from timestep to timestep.
Panel c in Fig.~3 in the main text shows a heatmap of energy error versus Krylov dimension and $dt$ for an example of our Hamiltonian on $20$ qubits.
In our experiments, we used heuristically chosen values of $dt$.

\subsection{Measurement bases}
\label{app:measurement_bases}


In this section we show that in one of our $k$-particle experiments, the measured observables can be partitioned into $2(k+2)$ co-measurable sets, or measurement bases.
These sets are shown in Table~\ref{tab:meas_bases}, and may be derived using the following identities:
\begin{equation}
\label{cnot_identities}
\begin{split}
    \mathrm{CX}_{01}\cdot X_0X_1\cdot\mathrm{CX}_{01}&=X_0I_1,\\
    \mathrm{CX}_{01}\cdot X_0I_1\cdot\mathrm{CX}_{01}&=X_0X_1,\\
    \mathrm{CX}_{01}\cdot Y_0I_1\cdot\mathrm{CX}_{01}&=Y_0X_1,\\
    \mathrm{CX}_{01}\cdot Y_0X_1\cdot\mathrm{CX}_{01}&=Y_0I_1,\\
    \mathrm{CX}_{01}\cdot X_0Z_1\cdot\mathrm{CX}_{01}&=-Y_0Y_1,\\
    \mathrm{CX}_{01}\cdot Y_0Z_1\cdot\mathrm{CX}_{01}&=X_0Y_1.
\end{split}
\end{equation}
Table~\ref{tab:meas_bases} is justified by \eqref{cnot_identities} together with that fact that, regardless of its circuit implementation, the controlled-initialization circuit is logically equivalent to the set of $\mathrm{CX}$s from the control qubit to each particle.
Also, we have utilized the symmetry to reduce the number of measurement bases. Namely, since the generator of the U(1) symmetry is $\sum_i Z_i$, and also the expectation values of all $XX$ and $YY$ terms of initial state (with $k$ particles) are equally zero, the $XX$ and $YY$ terms are interchangeable even after the Trotter evolution. This allows us to skip the measurement of $YY$ terms and simply focus on $XX$ terms.

For example, to explain the final row in Table~\ref{tab:meas_bases}, one notes that in the case of $Y$ on the control qubit and a $ZZ$ Hamiltonian term with one $Z$ on a particle, the controlled-initialization circuit contains a $\mathrm{CX}$ from the control to that particle: by the last identity in \eqref{cnot_identities}, the $YZ$ is transformed to $XY$ with the $X$ on the control and the $Y$ on the particle.
The remaining $\mathrm{CX}$s in the controlled-initialization circuit are then from the control to qubits not covered by this term, so the second identity in \eqref{cnot_identities} applies and places $Xs$ on the remaining particles.
Each $ZZ$ term with a $Z$ on this particular particle will be covered by this basis provided we measure all the remaining qubits in the $Z$ basis.

\begin{table*}[]
    \centering
    \begin{tabular}{c|c|c}
        Measurement basis type \hspace{0.2in} & \hspace{0.2in} $\#$ of bases \hspace{0.2in} & \hspace{0.2in} control qubit$\otimes$Hamiltonian term type \\\hline
        all $X$ & 1 & $X\otimes XX$ \\
        $Y$ on control; $X$ elsewhere & 1 & $Y\otimes XX$ \\
        $X$ on control, particles; $Z$ elsewhere & 1 & $X\otimes ZZ$ with $ZZ$ not on particles \\
        $Y$ on control; $X$ on particles; $Z$ elsewhere & 1 & $Y\otimes ZZ$ with $ZZ$ not on particles \\
        $Y$ on control; $Y$ on one particles, $X$ on others; $Z$ elsewhere & $k$ & $X\otimes ZZ$ with one $Z$ on a particle \\
        $X$ on control; $Y$ on one particle, $X$ on others; $Z$ elsewhere & $k$ & $Y\otimes ZZ$ with one $Z$ on a particle \\
    \end{tabular}
    \caption{Measurement bases when conjugating native Hamiltonian terms by controlled-initialization circuit. The left column describes the measurement basis, while the right column describes the native terms that the measurement basis corresponds to. This set of measurement bases assumes that none of the particles are on adjacent qubits, which guarantees that every $ZZ$ interaction in the Hamiltonian has at most one qubit on an particle. The assumption always holds below half-filling as long as the particles are distributed roughly uniformly over the qubit graph, as discussed in \cref{app:initialization}. This set of measurement bases also assumes that we do not need to explicitly measure the $YY$ terms in the Hamiltonian, since they will have the same values as the $XX$ terms by symmetry, as discussed in \cref{app:measurement_bases}. The last two rows each correspond to $k$ bases, one for each location of the $Y$ on one of the $k$ particles.}
    \label{tab:meas_bases}
\end{table*}

\subsection{Complete circuits}

\begin{figure}[t]
    \centering
    \vspace{-0.5in}
    \includegraphics[width=0.8\columnwidth]{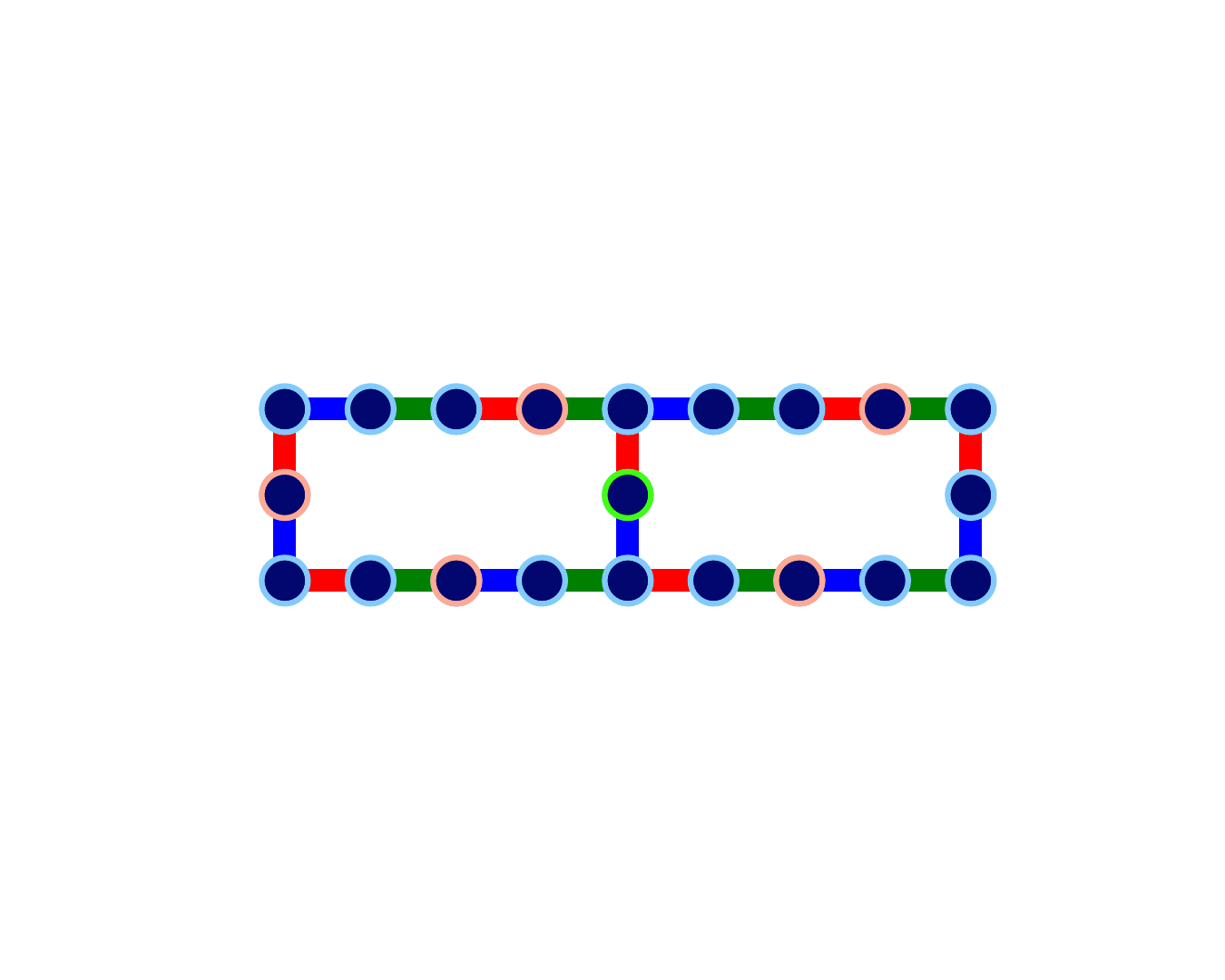}
    \vspace{-1in}\\
    \includegraphics[width=\columnwidth]{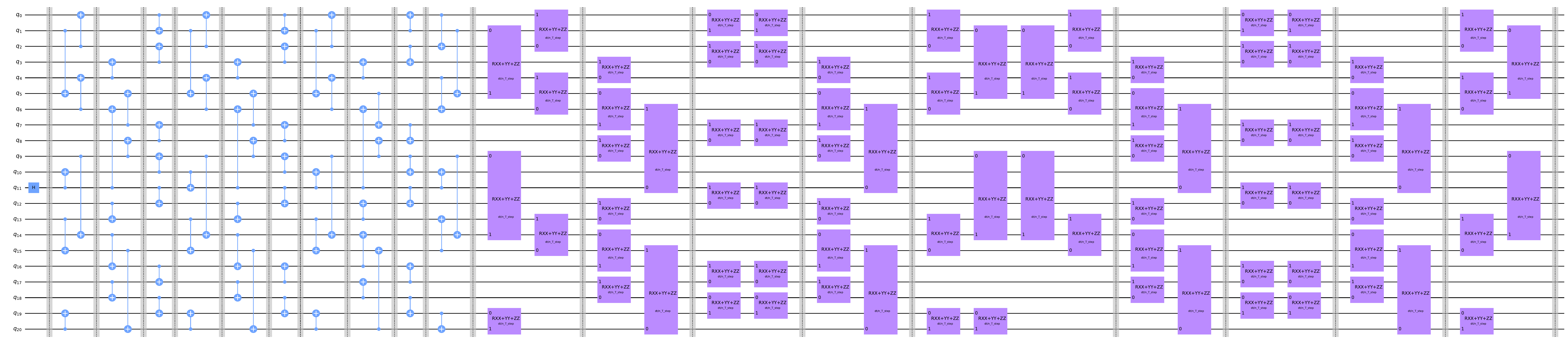}
    \caption{Example of the full quantum circuit for our algorithm, corresponding to the 21 qubit layout shown above. In the qubit layout, the green qubit is the control, the red qubits are the initial particle locations, and the edge colorings correspond to the simultaneously-implementable layers of two-qubit gates. The layers of two qubit gates (separated by barriers) in the circuit correspond to this edge coloring: transpilation and twirling transforms each $\mathrm{CX}$ in the controlled preparation (the first part of the circuit) into a CZ, and each purple two-qubit rotation in the Trotterized evolution (the second part of the circuit) into three CZs along with single-qubit gates.}
    \label{fig:full_circ_example}
\end{figure}

\cref{fig:full_circ_example} shows an example of a complete circuit.
Details are given in the caption.

\section{Device details}
\label{app:dev_details}

\paragraph*{Device overview.}
The experimental results shown are obtained by executing the quantum circuits of the Krylov quantum diagonalization algorithm on a Heron R1 device with 133 data qubits. Heron-type devices have fixed frequency transmon qubits~\cite{koch2007charge} as data qubits and tunable couplers. The tunable coupling allows for much faster interaction between physical qubits, reducing the two-qubit gate duration to be of the same order of the single-qubit gate duration. The reduction in duration of the two-qubit gates has important consequences in terms of gate errors and the type of error suppression and mitigation techniques that are more effective. 

\begin{figure}[t]
    \centering
    \includegraphics{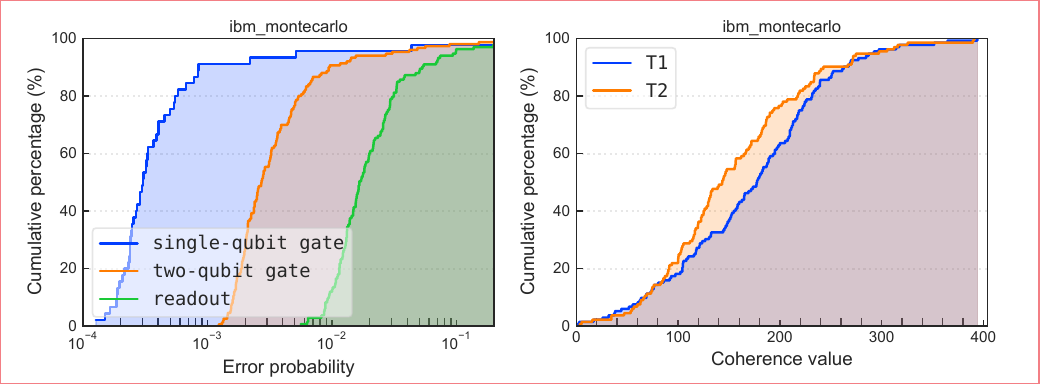}
    \caption{Device properties on June 16$^{\text{th}}$, 2024. The left panel shows cumulative percentages for various error rates, such as single-qubit gate error, two-qubit gate error, and readout error. The right panel shows the cumulative percentages of $T_1$ and $T_2$ coherence times in microseconds.}
    \label{fig:device-props}
\end{figure}

\paragraph*{Device properties.}
The IBM Quantum Heron processor's error properties are reported in Fig.~\ref{fig:device-props}. In the left panel, the single-qubit gate error, characterized by the randomized benchmarking technique, shows a mean error rate of $2.4 \times 10^{-2}$ with a significantly lower median value of $3.0 \times 10^{-4}$. This discrepancy between mean and median indicates a skewed distribution with a prevalence of lower error rates but occasional high outliers. The two-qubit gate error, also assessed via randomized benchmarking, has a mean error rate of $2.0 \times 10^{-2}$ and a median of $2.7 \times 10^{-3}$, reflecting a similar trend of generally low error rates interrupted by sporadic higher values.

The readout error, representing the readout assignment infidelity, has a mean of $3.6 \times 10^{-2}$ and a median of $1.6 \times 10^{-2}$. This substantial difference again points to an asymmetric distribution where most readout errors are relatively low, but there are instances of significantly higher infidelity.

The right panel of Fig.~\ref{fig:device-props} focuses on coherence times, important for the error rates in our experiment. The $T_1$ relaxation time, the period a qubit takes to relax to its ground state, shows a mean value of $1.7 \times 10^{2} \mu s$ and a median of $1.8 \times 10^{2} \mu s$, indicating a relatively stable and consistent performance with minimal variance. Similarly, the $T_2$ dephasing time, measuring the time over which a qubit maintains its quantum state coherence, has a mean of $1.5 \times 10^{2}\mu s$ and a median of $1.4 \times 10^{2}\mu s$, suggesting that dephasing times are slightly less stable than relaxation times but still within an acceptable range. These coherence times are indicative of the robust performance.

\begin{figure}[]
    \centering
    \includegraphics{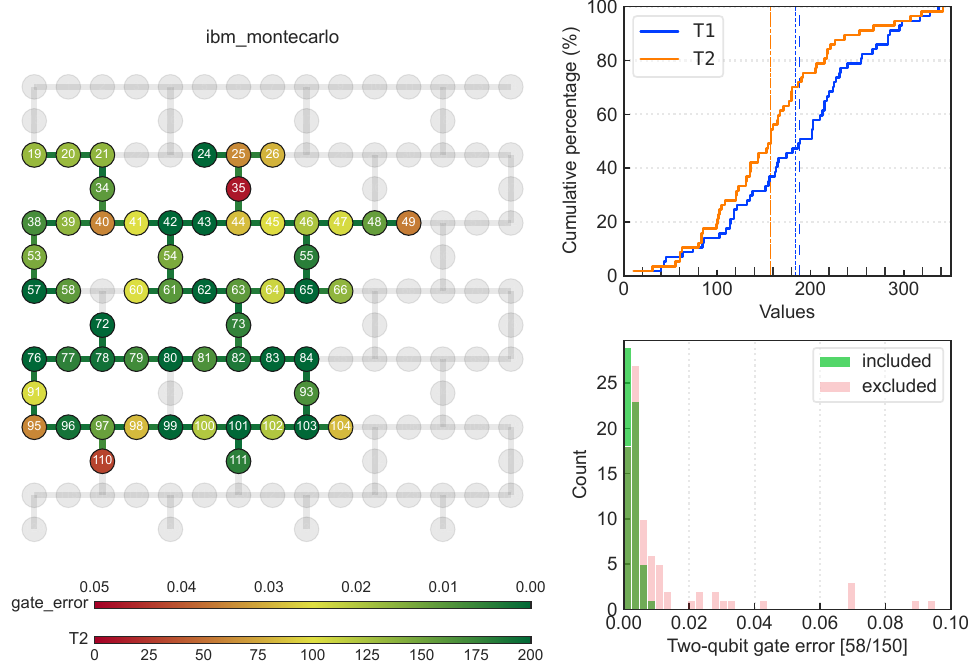}
    \caption{Qubit properties for the qubit subset selected for the $k=1$ experiment. The device map highlights two-qubit gate errors and $T_2$ coherence times. The panels on the right provide further information about $T_1$ and $T_2$ values, along with their mean (dotted)/median (dashed), as well as the distributions of two-qubit gate errors for the edges in the qubit subset chosen for the experiment compared to the edges excluded from it.}
    \label{fig:qubit_subset_k1}
\end{figure}

\begin{figure}[]
    \centering
    \includegraphics{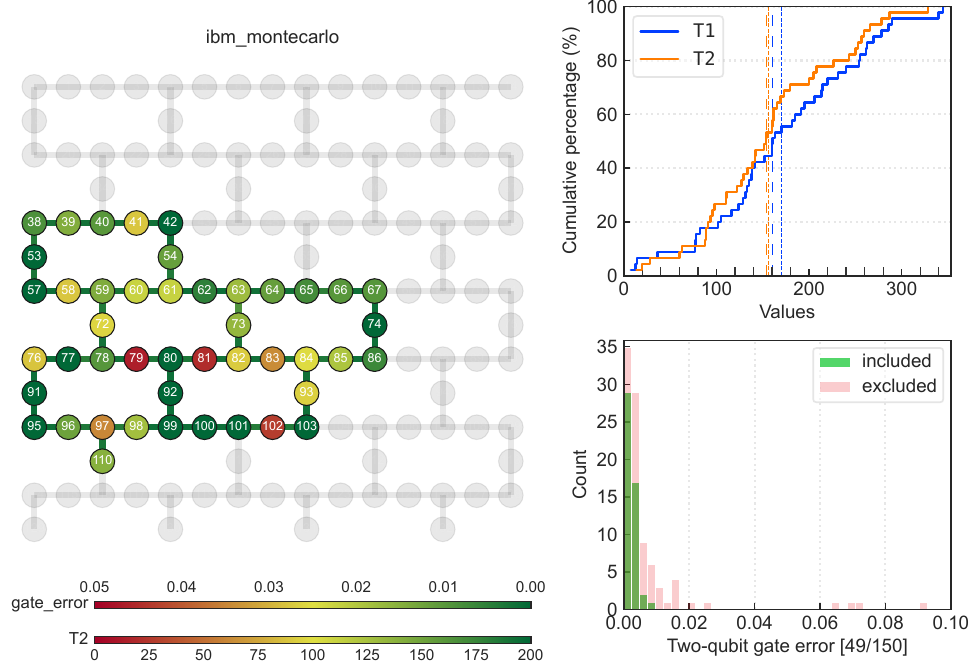}
    \caption{Qubit properties for the qubit subset selected for the $k=3$ experiment.}
    \label{fig:qubit_subset_k3}
\end{figure}

\begin{figure}[]
    \centering
    \includegraphics{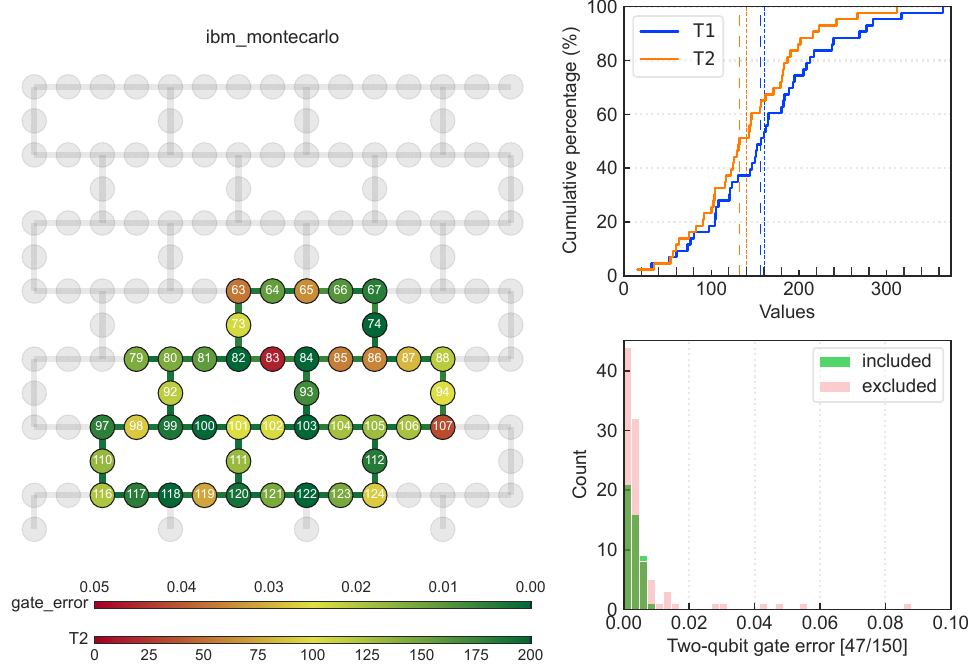}
    \caption{Qubit properties for the qubit subset selected for the $k=5$ experiment.}
    \label{fig:qubit_subset_k5}
\end{figure}

\paragraph*{Qubit used in the experiments.}
We reported experiments on different qubit subsets, which were chosen as the largest best performing subset at the time of the experiment. The 57 qubit subset selected for $k=1$ is shown in \cref{fig:qubit_subset_k1}. The qubits in this subset had the following median properties: $T_1 =180 \mu s$, $T_2=150 \mu s$, readout error $1.6\%$ and length $2 \mu s$, single-qubit gate error $0.022 \%$ and length $32 ns$, two-qubit gate error $0.24\%$ and length $104 ns$. The 45 qubit subset used for $k=3$ in \cref{fig:qubit_subset_k3}, had the following median properties: $T_1 =170 \mu s$, $T_2=140 \mu s$, readout error $1.6\%$ and length $2 \mu s$, single-qubit gate error $0.024 \%$ and length $32 ns$, two-qubit gate error $0.22\%$ and length $88 ns$.The 43 qubit subset used in the $k=5$ experiment \cref{fig:qubit_subset_k5} had the following median properties: $T_1 =160 \mu s$, $T_2=130 \mu s$, readout error $1.8\%$ and length $2 \mu s$, single-qubit gate error $0.027 \%$ and length $32 ns$, two-qubit gate error $0.25\%$ and length $88 ns$. Error rates for single and two-qubit gates are calculated via randomized benchmarking while the readout error is calculated as the average rate of mis-classification of the 0/1 state when the 1/0 state is prepared.




\section{Error suppression and mitigation}
\label{app:err_supp}

\paragraph*{Error suppression and mitigation workflow.}
To reduce the error in the estimated expectation values, we employ a series of error suppression and mitigation techniques. Error suppression techniques have little or no cost in terms in resource overhead while error mitigation techniques typically involve an exponential increase in classical or quantum resources~\cite{tsubouchi2023universal, takagi2023universal}. These become useful when the resulting exponential increase in resources is favorable compared to the exponential cost with the number of qubits of a full classical simulation of the circuits. The error suppression techniques used in the experiments consist of a heuristic method for layout selection (Sec.~\ref{sec:layout_sel}) and the insertion of dynamical decoupling (Sec.~\ref{sec:dyn_dec}) sequences wherever qubits are idling for a sufficiently long time (to accommodate the dynamical decoupling sequence) in the circuit.

In addition, we use several error mitigation techniques that can be combined together: Pauli twirling (Sec.~\ref{sec:pauli_twirl}), measurement error mitigation (Sec.~\ref{sec:trex_mit}) and Probabilistic Error Amplification (PEA~--- Sec.~\ref{sec:PEA}). As will become apparent from the details of the error mitigation and suppression techniques described below, a key component of the pipeline is the tailoring of the noise into Pauli noise that is enabled by Pauli twirling. For this reason we briefly review the efficient characterization of this noise using sparse Pauli-Lindblad noise models in Sec.~\ref{sec:SPL}.

\subsection{Dynamical decoupling}
\label{sec:dyn_dec}
Dynamical decoupling is a technique that aims at removing the contribution of unwanted interaction terms in the Hamiltonian that determines the time-evolution of the system of interest~\cite{hahn1950spin, viola1999dynamical, khodjasteh2005fault, ezzell2023dynamical, seif2024suppressing}. It is considered an open-loop control technique as it does not involve measuring and incorporating the information of the effects generated by the application of the control action. In quantum systems, it can be used to remove coherent and incoherent evolutions. Its uses range from decoherence suppression to averaging out couplings between system and environment, halting the natural evolution of the system to retain the information in the quantum state. Ultimately, the effects that can be suppressed with dynamical decoupling sequences depend on the difference in time-scales between the correlation times of these unwanted effects and the minimum accessible control time-scale. 

In the context of the experiment considered here, dynamical decoupling sequences are inserted during idle times of a qubit in a quantum circuit to reduce the effects of cross-talk due to the control of neighboring qubits. Ref.~\cite{seif2024suppressing} shows different examples of dynamical decoupling sequences that remove the main noise contributions for superconducting qubit devices. For Eagle type devices, which have fixed always-on coupling between qubits, it is important to reduce error generated by cross-talk between qubits. However, the tunable coupling characteristic of Heron type devices, the one used in this demonstration, reduces the effects of crosstalk and thus the need to suppress these contributions. In the experiment, we opt for an $X +, \; X-$ dynamical decoupling sequence, where the $\pm$ sign indicates that we take $X$ pulses with opposite amplitude. This type of sequence removes single-qubit Z errors that happen during the idling time of the qubits and corrects for any imperfections in the amplitude of the X pulse.

\subsection{Twirling}
\label{sec:pauli_twirl}

Twirling is the operation of averaging a noise channel $\Lambda$ by conjugating it with a set of unitary channels $\mathcal{U}$ 
according to some measure $\eta$ over a subgroup $\mathcal{X}$ of the unitary group $\mathscr{U}$:  

\begin{equation}
\label{eq:twirl}
    \Lambda \longrightarrow \int_{\mathcal{U} \in \mathcal{X}} d\eta \; \mathcal{U}^{\dagger} \circ \Lambda \circ \mathcal{U}.
\end{equation} 

\noindent
Depending on the choice of measure $\eta$ and the subgroup $\mathcal{X}$, the effect of twirling varies. For example, in the case where $\eta=\mu_{Haar}$ is the Haar measure and the subgroup is identical to the entire unitary group as $\mathcal{X}=\mathscr{U}$,
a generic noise channel $\Lambda$ can be twirled into a global depolarizing noise channel. 

By introducing the idea of unitary designs~\cite{dinitz2007handbook}, we can turn the integration into a sum over a finite number of elements. The unitary design is defined as follows:
assume that we are interested in average quantities of a polynomial of degree $t$ over a uniform distribution of a certain set $\mathcal{X}$. Then, if the average of function evaluation over a uniform distribution from a subset $\mathcal{X}$ yields identical value to that of the Haar integral over the entire group, we say that such a subset constitutes a $t$-design. In particular, a unitary $t$-design refers to a subset of the unitary group that mimics the integration over the Haar-random distribution up to the $t$-th moment.
It can be shown~\cite{dankert2009exact}, that a unitary 2-design is sufficient to twirl a noisy channel.
For multiqubit systems, one may in principle use the Clifford group (which actually forms a unitary 3-design) to twirl a noisy channel and obtain a global depolarizing noise channel, although for current devices this is not a compelling option since an $N$-qubit Clifford operator requires $O(N)$ depth in general~\cite{bravyi2021hadamard}. 
Additionally, the fact that the noise channel is intertwined with the corresponding unitary operation means that twirling gates must be commuted through the unitary operation that is being twirled; in other words, non-Clifford gates cannot be twirled by the Clifford group without use of additional non-Clifford gates. 

This has led to the idea of compromising on the twirled output channel: instead of requiring that it be the global depolarizing channel, one can aim for a less homogeneous but still analyzable noise channel by taking different groups for the twirling operation.
An important example is given by the Pauli group: randomly choosing unitaries from the Pauli group leads to Pauli twirling. The effect of Pauli twirling is to turn a noise channel into a Pauli noise channel, which is a diagonal matrix in the Pauli Transfer Matrix (PTM) representation~\cite{dankert2009exact}. Throughout the experiments presented in this paper, Pauli twirling was employed for learning and mitigating noise channels associated with layers of two-qubit gates and measurements~\cite{vandenberg2023probabilistic,van2022model}.

\subsection{Sparse Pauli-Lindblad noise model}
\label{sec:SPL}
The Pauli-Lindblad noise model is a framework introduced in Ref.~\cite{vandenberg2023probabilistic} used to describe noise channels, particularly under the assumption of locally correlated noise. This model is grounded in the continuous-time Markovian dynamics represented by the Lindblad master equation, \( \frac{d\rho(t)}{dt} = \mathcal{L}(\rho(t)) \), which gives a phenomenological way to describe the evolution of the system's density matrix in the presence of noise. Here, $\mathcal{L}$ is a superoperator called a Lindbladian.
The Pauli-Lindblad model focuses on the contributions from Pauli operators, eliminating the internal Hamiltonian dynamics to simplify the noise representation. The model constructs Lindblad operators \( A_k \) as linear combinations of Pauli operators \( P_k \), characterized by a set of non-negative coefficients \( \lambda_k \), such that \( A_k = \sqrt{\lambda_k} P_k \). This approach facilitates a sparse representation of the noise. Concretely, by mapping the noise channel superoperators $\Lambda$ into operators $\hat{\Lambda}$ by using the  the Choi-Jamiolkowski isomorphism, we can express the noise as \( \hat{\Lambda} = e^{\hat{\mathcal{L}}} \), where \( {\hat{\mathcal{L}} = \sum_{k \in K} \lambda_k (P_k \otimes P_k^T - I \otimes I)} \), and the set of Paulis $K$ has size $\vert K \vert$ polynomial in the number of qubits, $\vert K \vert \ll 4^n -1$. This is motivated by the locality of the noise due to the sparse connectivity of physical qubits on the device. The resulting noise model can be expressed in terms of matrix exponentials, \( \hat{\Lambda} = \prod_k e^{-\lambda_k P_k \otimes P_k^T} \), which allows for efficient characterization of the parameters of the noise model. For example, protocols like cycle benchmarking~\cite{erhard2019characterizing} can be used to characterize the fidelity \( f_a \) of a Pauli operator \( P_a \), given by \( f_a = \frac{1}{2^n} \text{Tr}[P_a^\dagger \Lambda(P_a)] \), which in turn gives us the information we need to describe the model parameters.  

\subsection{Layout selection}
\label{sec:layout_sel}

The virtual-to-physical qubit mapping problem, also known as qubit mapping or layout selection, involves optimally assigning virtual qubits of a quantum circuit to physical qubits on a quantum processor to extract the best algorithmic performance. This is achieved by selecting qubits and gates with lowest error rates while optimizing gate connectivity. To execute a virtual circuit on a device with different connectivity, SWAP operations are inserted to implement the circuit's interactions on the device's topology. This procedure is commonly known as routing. Routed circuits using a subset of device qubits can be mapped to different regions without extra routing operations, requiring the selection of a subset of qubits with optimal performance, which is the core of the layout selection problem. Mapping methods typically use data from characterization/calibration experiments for selecting qubit layouts for a circuit. One of the most popular methods, mapomatic~\cite{nation2023suppressing}, finds qubit subsets in the device topology matching a circuit's graph and uses device calibration data to score the different qubit subsets. In this experiment we have employed two methods for selecting the qubits on which to run the experiment. 


For the experiment with a single particle ($k=1$), which involves a relatively shallow controlled-initialization subcircuit (only 1 $\mathrm{CX}$ from the control qubit to the chosen excited qubit), we have used the MAESTRO~\cite{amico24maestro} mapping method to choose the largest subset of qubits (without restriction to its topology) with some desired properties. The chosen subset, along with the qubit properties is shown in \cref{fig:qubit_subset_k1}. First, MAESTRO uses the information gathered from the learning of a sparse Pauli-Lindblad noise model for the whole device. It does so by partitioning the topology of the device into the minimum number colorings needed to cover the entire set of edges of the device. In the case of the Heavy-hex lattice, the set of edges can be partitioned into three set of non-overlapping edges or a 3-coloring of the graph as shown in \cref{fig:edge_coloring_example}. Then MAESTRO employs the same techniques used in PEC/PEA~\cite{vandenberg2023probabilistic, kim2023evidence} to learn a sparse representation of the noise channel of the three layers of two-qubit gates corresponding to the 3-coloring and the SPAM error of the qubits. Once these have been learned, the error generators are marginalized (thus neglecting cross-talk) to describe only the noise models for each of the edges. This procedure results in a compact representation of the noise in the two-qubit gates (single-qubit gates are assumed to be noiseless) and in the measurement.

Finally, to quantify the quality of a subset of qubit of a certain size, the circuit to execute is considered.
MAESTRO (like mapomatic) finds the subset of qubits within the device topology that have the same size and interaction graph of the circuit. That is, the edges in the qubit subset allow the two-qubit interactions in the circuit to be mapped directly to the physical qubits without extra routing. Each of the compatible subsets is scored using a cost function that takes into account the error generators of the two-qubit gates present in the circuit, and the measured qubits. The subset with the best (lowest) score is chosen.

For the $k=1$ experiment, varying circuit widths were evaluated to select the largest subset of qubits with a score below a certain threshold. The downside of this approach is that we cannot impose any requirement on the topology of the selected subset, which may be of any type as long as it is compatible with the circuit. For higher numbers of particles, $k=3, 5$, which require spreading entanglement across many of the qubits in the layout, the chosen subgraph of the heavy-hexagonal topology additionally determines the depth of the controlled-initialization subcircuit, with inclusion of more complete heavy-hexes being favorable in particular. The MAESTRO tool did not allow for optimizing with respect to this additional figure of merit, so at higher numbers of particles the qubit subsets were chosen by hand selecting the largest subsets of qubits in the heavy-hexagonal device lattice such that the corresponding Hamiltonian interaction lattices contained complete heavy-hexes and the qubit properties ($T_1$, $T_2$, $Err_{2Q}$, $Err_{1Q}$, SPAM) met certain thresholds. Selected subsets are shown in \cref{fig:qubit_subset_k3,fig:qubit_subset_k5}.

\subsection{Measurement error mitigation}
\label{sec:trex_mit}
Measurement error mitigation refers to techniques used to reduce inaccuracies in the measurement outcomes of quantum processors caused by hardware imperfections and noise, which introduce bias into the expectation values of observables. We consider the measurement error mitigation technique introduced in Ref.~\cite{van2022model}, which is a scalable method for addressing these errors without requiring  detailed noise models. The method applies random Pauli bit flips $X$ to qubits before measurement, transforming the noise into a measurable multiplicative factor $\lambda$. Assuming that we measure in the Z basis, the noisy expectation value becomes $\langle \tilde{Z} \rangle^{\ast} = \lambda \langle Z \rangle$, where $\langle Z \rangle$ denotes the expectation value without readout errors. This effect is obtained when averaging over multiple circuit instances, each with different random bit flips. The method leverages the diagonalization of the noise channel, which is achieved when twirling the noise channel as seen in Sec. \ref{sec:pauli_twirl}. If we only care about Z expectation values, this is obtained by applying random Pauli bit flips $X$ and averaging. 
To mitigate twirled expectation values, we first need to measure the multiplicative factor $\lambda$ with benchmark circuits and then rescale the twirled expectation values of interest by the inverse of that factor. This approach corrects bias in expectation values efficiently, even for large quantum systems, and effectively mitigates correlated readout errors with minimal additional sampling complexity. The diagonal elements $\lambda$ can be measured directly, allowing for the mitigation of readout errors without requiring an explicit noise model.

\subsection{Probabilistic error amplification}
\label{sec:PEA}

Probabilistic Error Amplification (PEA) is a technique introduced in Ref.~\cite{kim2023evidence} to mitigate errors in quantum computations by learning the noise of the system and amplifying at different strengths to extrapolate results back to the ideal, zero-noise limit. Unlike previous implementations of Zero Noise Extrapolation (ZNE)~\cite{temme2017error}, this approach leverages a sparse Pauli-Lindblad noise model~\cite{vandenberg2023probabilistic} for the noise amplification. As described in Sec.~\ref{sec:SPL}, the noise in each layer of two-qubit gates is modeled by $\mathcal{L}(\rho) = \sum_{k \in K} \lambda_k (P_k \rho P_k^\dagger - \rho)$, with $\lambda_k$ representing the noise rates and $P_k$ being Pauli operators. To amplify the noise, the error rates are scaled by a factor $\alpha$, resulting in $\mathcal{L}_\alpha(\rho) =  \sum_{k \in K} \alpha  \lambda_k (P_k \rho P_k^\dagger - \rho)$.

The practical implementation of PEA involves learning the noise model through characterization experiments, where the probabilities of various Pauli errors are estimated, and mitigating the measured results by executing the circuits at different noise levels and extrapolating the measured expectation values to the zero-noise limit. Pauli twirling is employed to simplify the noise into a Pauli noise channel, making it easier to measure and model its error rates (see Sec.~\ref{sec:pauli_twirl}). 
For the mitigation, single-qubit Pauli gates are inserted before each Pauli-twirled layer of two-qubit interactions (e.g. CXs) to allow for the insertion of noise amplified at a certain gain $G$. The noise gain $G$ is related to the amplification of the noise channel by a factor $\alpha$ where $G=\alpha + 1$, since the introduction of extra noise as a noise channel $\Lambda^{\alpha}= e^{\alpha \mathcal{L}}$ on top of the native noise channel $\Lambda$ results in a total noise channel $\Lambda^G = \Lambda^{\alpha + 1}$. In the case of noise amplification, one can directly sample these extra single-qubit Pauli operators to implement the desired noise channel. For each level of amplified noise, the expectation values of observables of interest are measured. These noisy results are then used to extrapolate the ideal values at zero noise using either linear or exponential extrapolation methods. The motivation behind this choice is expressed in Sec. V of the supplementary material in Ref.~\cite{kim2023evidence}. Linear extrapolation assumes a simple linear relationship: $\langle O \rangle_G \approx a + bG $, where the parameters $a,b$ are determined by fitting a line through the measured values of $\langle O \rangle_G$ at different noise levels $G$ and the noise-free value corresponds to $\langle O \rangle_0=a$. Exponential extrapolation models the decay as $\langle O \rangle_G \approx ae^{-bG}$, which again gives $\langle O \rangle_0=a$.

In our experiments, for each measurement basis a certain number of twirled instances were generated and each instance was then repeatedly measured, for different values of the noise amplification factor. For the single-particle ($k=1$) experiment, we used 300 twirled instances with 500 shots each, at noise amplification factors of $1, 1.5, 3$. For $k=3,5$, we used 100 twirled instances with 500 shots each, at noise factors $1, 1.3, 1.6$.

The reduction in twirled instances for the larger experiments was introduced in order to reduce the total runtime, since the numbers of measurement bases as well as the circuit sizes increase with $k$ as discussed above. The adjustment of the noise amplification factors was due to the increased noise rates in the deeper circuits. The controlled-initialization part of the circuit involves creating a maximally entangled state of the control qubit and the initial particle locations. With an increase in the number of particles, this translates to a larger maximally entangled state prepared at the beginning of the circuit, which in turns makes the results more susceptible to noise.

\subsubsection{Extrapolation criteria}
\label{sec:fit}
We have developed an automated algorithm for choosing the most appropriate extrapolation method for each of the observables. Ideally, we would always use the exponential fit. However, in the case where the expectation values at different noise levels are too close to zero, the exponential fit can become unstable and yield diverging values for the extrapolated expectation value. This can happen for both in the situation where the noise-free expectation value is zero and when the noise is too strong to measure any signal for a particular observable. In these cases, a linear fit would be preferred. 

To avoid manually deciding between fit methods, we automate the downgrading from an exponential to a linear fit by checking if any of the following criteria fails:
\begin{itemize}
    \item Expectation value is confined: the extrapolated expectation value for the Pauli observable is contained within the interval $[-1, \; 1]$.
    \item Exponential fit is a better fit: the calculated $\chi^2_{\exp}$ of the exponential fit is lower than the $\chi^2_{lin}$ of the linear fit.
    \item Extrapolation error is low for the exponential fit: the ratio between the standard deviation of the extrapolated expectation value and the extrapolated expectation value itself is lower than $0.5$.
\end{itemize}

\section{Post-processing}
\label{app:post_processing}

\subsection{Automated regularization}
\label{sec:automated_reg}

As mentioned above in \cref{app:overview}, we used a heuristic to automate the choice of threshold $\epsilon$ used to regularize the generalized eigenvalue problem \eqref{eq:gen_eig_prob_app} in the classical post-processing.
There were three main reasons this was necessary: (i) the theoretically-optimal choice~\cite{epperly2021subspacediagonalization,kirby2023exactefficient} was not available to us since we did not have a sufficiently precise quantification of the noise in $(\widetilde{H},\widetilde{S})$, (ii) to eliminate bias we might have introduced by choosing thresholds by hand, and (iii) because we also employed bootstrapping to compute error bars, and regularizing each bootstrapped run would have been impractical.

The heuristic we developed exploits theoretical knowledge of the behavior of the output energy estimates as a function of the Krylov dimension~\cite{epperly2021subspacediagonalization,kirby2023exactefficient}.
As discussed in the main text, one key takeaway from these results is that even in the presence of noise, a well-conditioned energy versus Krylov dimension curve is an exponential decay, just with a rate and asymptote that depend on the noise.
However, poorly conditioned curves instead tend to show large fluctuations away from the smooth exponential decay, while if noise has completely dominated the signal, the convergence tends to be approximately linear with a small negative slope.
Therefore, our heuristic was as follows:
\begin{enumerate}
    \item Set an initial threshold $\epsilon$ much smaller than the noise rate. We used $10^{-8}\cdot D$ (see below for an explanation of the factor $D$), which would be orders of magnitude smaller than our noise rate even if the only source of noise were finite sampling.

    \item Perform a logarithmic search over increasing thresholds to find the smallest threshold for which the resulting energy versus Krylov dimension curve fits an exponential decay up to a chosen tolerance.

    \item Return the final threshold and the corresponding energy versus Krylov dimension curve.
\end{enumerate}
We chose the tolerance on the quality of the fit to be $0.5\cdot D$ in rms error (i.e., an average deviation of $0.5$ at each point) by hand using preliminary results for the single-particle experiment, choosing the value that led to good performance.
In order not to bias our results, we then used the same tolerance in all of our main experiments regardless of number of particles, qubit number, and layout.
A final technical detail is that the theory~\cite{epperly2021subspacediagonalization,kirby2024analysis} indicates that the threshold should grow with the Krylov dimension: in the presence of i.i.d. Gaussian noise on the matrix elements, the rate may be as low as $\widetilde{O}(\sqrt{D})$~\cite{lee2023sampling}, but for generic noise the rate is $O(D)$.
Since we wished to avoid as much as possible any assumptions on the noise in our matrices $(\widetilde{H},\widetilde{S})$, we used the latter scaling: hence the factor of $D$ in the thresholds.

The resulting thresholds were $10^{-8}$, 0.11, and $10^{-8}\cdot D$ for $k=1,3,5$, respectively.
The $10^{-8}\cdot D$ for $k=1$ and $5$ indicates that the heuristic found a good fits to an exponential decay at its initial threshold guess, and hence did not proceed further.
While this might appear surprising, it is really just because the positive eigenvalues of $\widetilde{S}$ were well-separated from zero in these two experiments, and the heuristic search begins from a low threshold and searches upward.
To assess this, we found the lowest positive eigenvalue at each $D$ and divided each of those by the corresponding $D$: the minimum of the resulting values was $\sim0.0030$ for the $k=1$ experiment and $\sim0.0089$ for the $k=5$ experiment.
Hence increasing the threshold all the way up to those points would make no difference in the results, i.e., $0.0029$ and $0.0088$ would have been equivalent thresholds for $k=1$ and $k=5$, respectively.

\subsection{Bootstrapping}
\label{sec:bootstrapping}

We used bootstrapping to estimate the variances on the experimental results shown in Fig.~4 in the main text.
The bootstrapping was implemented at the shots level, so an entire single bootstrap involved the following steps:
\begin{enumerate}
    \item For each measurement basis and each twirl setting, resample a new set of measurement outcomes from the original set, with replacement.
    
    \item Calculate the corresponding expectation values, and from these the corresponding matrix elements of $(\widetilde{H},\widetilde{S})$, as in \cref{app:circuit_derivation}.
    
    \item Use the automated regularization heuristic described in \cref{sec:automated_reg} to choose a threshold for the bootstrapped matrix pair $(\widetilde{H},\widetilde{S})$, and solve the generalized eigenvalue problem \eqref{eq:gen_eig_prob_app} with that regularization threshold.

    \item Either accept or discard the resulting bootstrapped energy versus Krylov dimension curve, with the following criteria:
    \begin{enumerate}
        \item Reject if any of the energies at higher Krylov dimensions are above the energy of the initial reference state (i.e., the energy at $D=1$).

        \item Reject if the curve fit used by the automated regularization fails to converge at any point in the logarithmic threshold search.

        \item Otherwise accept.
    \end{enumerate}
    Note that both of these rejection criteria are well-motivated and do not require any special knowledge of the problem instance.
    The energy at a higher Krylov dimension exceeding the initial energy indicates that the effective noisy subspace is so poorly conditioned that the regularization more than cancels out the energy improvement with increasing dimension.
    And similarly, failure of the curve fit to converge indicates wildly fluctuating data.
\end{enumerate}

Once a sufficiently large collection of bootstrapped energy curves were accepted, those were used to calculate the standard deviations on the energies, which are shown in the main text in Fig.~4.
Some additional details of the bootstrapping are given in \cref{tab:bootstrapping_details}.

\begin{table}[t]
    \centering
    \begin{tabular}{c|c|c|c|c|c}
        $k$ & min $\epsilon$ & max $\epsilon$ & median $\epsilon$ & $\#$ bootstraps accepted & $\#$ bootstraps rejected \\\hline
        1 & $10^{-8}$ & $0.277$ & $0.015$ & 557 & 403 \\
        3 & $10^{-8}$ & $0.333$ & $0.144$ & 622 & 554 \\
        5 & $10^{-8}$ & $0.231$ & $10^{-8}$ & 1256 & 15
    \end{tabular}
    \caption{Additional bootstrapping details. $k$ is particle number (i.e., which experiment the row corresponds to). $\epsilon$ is the regularization threshold, which as described in \cref{app:post_processing}, was chosen automatically and independently for each bootstrapped matrix pair. The values of $\epsilon$ are given at $D=1$; as discussed in \cref{sec:automated_reg}, these are multiplied by $D$ to obtain the actual threshold at a given Krylov dimension. The minimum value of $\epsilon$ was $10^{-8}$, which was found to be adequate for 170 matrix pairs in the $k=1$ experiment and one in the $k=3$ experiment. This indicates that these bootstrapped pairs happened to come out well-conditioned; it is not surprising that this can happen occasionally, even when the original experimental matrix pair requires a higher threshold. On the other hand, $\epsilon=10^{-8}$ was adequate for 1171 of the 1256 accepted resamples for $k=5$: this may be attributed to the fact that it was also adequate for the original experimental matrix pair, i.e., the lowest energy state in the Krylov space was not too badly perturbed by the noise in this experiment (see \cref{sec:automated_reg} for further discussion). The final two columns indicate that in all experiments, approximately half of the bootstrapped matrix pairs were rejected as too ill-conditioned; see \cref{sec:bootstrapping} for details.}
    \label{tab:bootstrapping_details}
\end{table}

\end{document}